\documentclass[aps,prx,twocolumn,secnumarabic,amsmath,amssymb,superscriptaddress,longbibliography]{revtex4-1}

\AtBeginDocument{\usepackage{booktabs}}               
\usepackage{microtype}
\usepackage{wasysym}
\usepackage[varg]{txfonts}
\usepackage{hyperref}
\usepackage{amsmath}
\usepackage[percent]{overpic}
\usepackage{mathrsfs}
\usepackage{braket}
\usepackage{titlesec}
\usepackage{bm}
\usepackage{comment}
\usepackage{verbatim}
\usepackage{tabularx}
\usepackage{graphicx}
\usepackage{pict2e}
\usepackage{subfigure}
\usepackage{color}

\definecolor{darkred}{rgb}{0.7,0.0,0.0}
\def\cbl{\color{blue}}
\definecolor{darkblue}{rgb}{0,0.02,0.45}

\definecolor{darkgreen}{rgb}{0.02,0.45,0.0}

\definecolor{violet}{rgb}{0.8,0.2,0.6}

\definecolor{cL}{RGB}{59, 83, 140}
\definecolor{cM}{RGB}{33, 145, 141}
\definecolor{cH}{RGB}{95, 202, 98}
\definecolor{cG}{RGB}{204,204,204}

\def\be{\begin{equation}}
\def\ee{\end{equation}}
\def\bea{\begin{eqnarray}}
\def\eea{\end{eqnarray}}

\def\vec{\mathbf}
\def\bs{\boldsymbol}
\def\mc{\mathcal}

\hypersetup{colorlinks=true,linkcolor=red,citecolor=blue,urlcolor=blue} 
\usepackage[nolist,nohyperlinks]{acronym}

\begin{document}
\date{\today}

\title{Quantum spin liquid at finite temperature: proximate dynamics and persistent typicality}

\author{I. Rousochatzakis}
\affiliation{Department of Physics and Centre for Science and Materials, Loughborough University, Loughborough LE11 3TU, United Kingdom}\affiliation{School of Physics and Astronomy, University of Minnesota, Minneapolis, MN 55455, USA}
\author{S. Kourtis}\affiliation{Department of Physics, Boston University, Boston, MA, 02215, USA}
\author{J. Knolle}\affiliation{Blackett Laboratory, Imperial College London, London SW7 2AZ, United Kingdom}
\author{R. Moessner}\affiliation{Max Planck Institute for the Physics of Complex Systems, D-01187 Dresden, Germany}
\author{N. B. Perkins}\affiliation{School of Physics and Astronomy, University of Minnesota, Minneapolis, MN 55455, USA}

\begin{abstract}
Quantum spin liquids are long-range entangled states of matter with emergent gauge fields and fractionalized excitations. While candidate materials, {such as the Kitaev honeycomb ruthenate $\alpha$-RuCl$_3$}, show magnetic order at low temperatures $T$, here we demonstrate numerically a dynamical crossover from magnon-like behavior at low $T$ and frequencies $\omega$ to long-lived fractionalized {\it fermionic} quasiparticles at higher $T$ and $\omega$. This crossover is akin to the presence of spinon continua in quasi-1D spin chains. It is further shown to go hand in hand with persistent {\it typicality} down to very low $T$. This aspect, which has also been observed in the spin-1/2 kagome Heisenberg antiferromagnet, is a signature of proximate spin liquidity and emergent gauge degrees of freedom more generally, and can be the basis for the numerical study of many finite-$T$ properties of putative spin liquids. 
\end{abstract}

\maketitle

\pagebreak

\section{Introduction}\vspace*{-0.2cm}
Quantum spin liquids (QSLs) have been one of the central themes in condensed matter for many years.~\cite{Anderson1973,FazekasAnderson74,Kalmeyer1987,Wen1989,Moessner2001, Kitaev2003,Kitaev2006,Balents2010,Savary2016,Zhou2017,IoannisClassKitaev} Unlike conventional phases of matter, the characteristic correlations in QSLs are  non-local and cannot be detected directly by standard probes. As such, identifying the experimental signatures of QSLs is one of the most challenging tasks in the field.~\cite{Knolle2018} This task has become all the more pressing in the last decade, with the discovery of several candidate materials, including isotropic layered kagome systems~\cite{Balents2010,Shores05,Han2012,Norman2016} and the strong spin-orbit coupled iridates and ruthenates.~\cite{Jackeli2009,Jackeli2010,BookCao,Balents2014,Rau2016,Trebst2017,Hermanns2017,Winter2017, Singh2010, Singh2012,Liu2011,
Sears2015,Johnson2015,
Williams2016,
%
Biffin2014b, 
Modic2014,Biffin2014a,Takayama2015}

A promising route to detect QSLs is to look for signatures of fractionalization in dynamical probes, such as inelastic neutron scattering (INS),~\cite{Nagler91,Tennant93,Mourigal2013,Knolle2015,Knolle2014a,Banerjee2016,Banerjee2017}  Raman scattering,~\cite{Wulferding2010,Ko2010,Sandilands2015,Sandilands2016,Glamazda2016,Knolle2014,Nasu2016} resonant inelastic x-ray scattering (RIXS),~\cite{Gabor2016,Halasz2017,Savary2015} and ultrafast spectroscopy.~\cite{Alpichshev15} Such probes couple to multiple fractionalized quasiparticles, leading to characteristic broad scattering profiles. 
However, potential QSLs tend to be sensitive to perturbations~\cite{Jackeli2010,Schaffer2012,Jackeli2013,Lee2014,Katukuri2014,Katukuri2015,IoannisK1K2,Satoshi2016} and are in fact, in many if not most cases, preempted by magnetic order at low temperatures $T$.~\cite{
Singh2010, Singh2012,Liu2011,
Sears2015,Johnson2015,
Williams2016,
%
Biffin2014b, 
Modic2014,Biffin2014a,Takayama2015}

\begin{figure}[!b] 
\centering
\includegraphics[width=0.45\textwidth,angle=0,clip=true,trim=0 0 0 0]{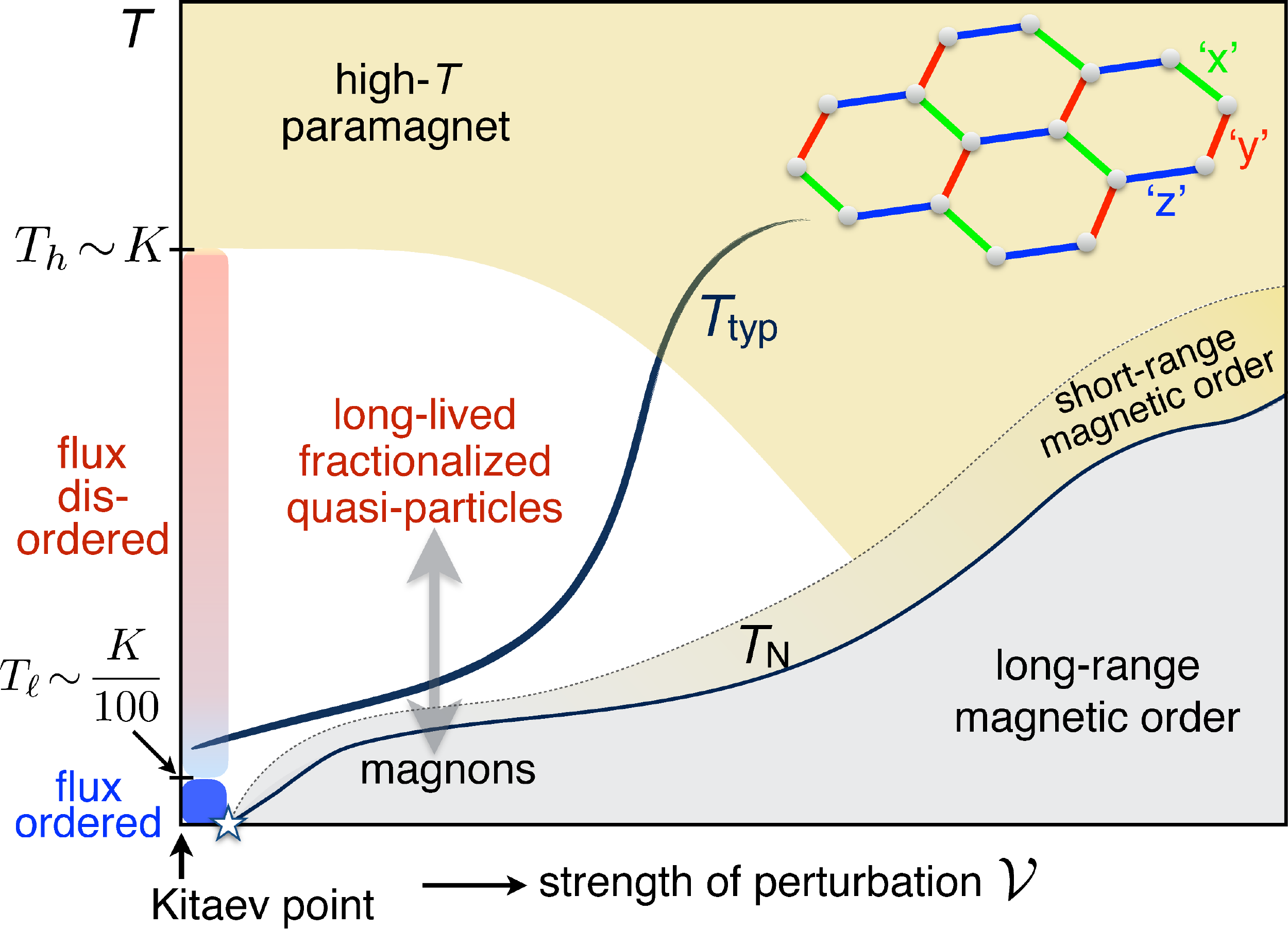}
\caption{{\bf Dynamical crossover near a Kitaev QSL (schematic).} Magnon excitations, characteristic of the long-range magnetic order at low $T$ and $\omega$, give way to long-lived fractionalized quasi-particles (here Majorana fermions) at higher $T$ and $\omega$. This crossover is accompanied by a remarkable persistence of {\it typicality} to very low $T$ near the QSL.}\label{fig:Cartoon}
\end{figure}

Despite this, here we establish numerically that long-lived fractionalized quasi-particles are still present in the spectrum at finite energies, in the same way that spinon continua survive in quasi-1D spin chains, which also order at low $T$ due to weak interchain interactions.~\cite{Nagler91,Tennant93,Zheludev2000,Zaliznyak2004,Lake2005,Enderle2010,Mourigal2013,Bera2018}  
Fig.~\ref{fig:Cartoon} illustrates the qualitative picture for the case of honeycomb Kitaev materials, like $\alpha$-RuCl$_3$, which are proximate to a gapless spin liquid characterized by emergent magnetic fluxes and Majorana fermions.~\cite{Kitaev2006} Here, the characteristic ordering temperature $T_N$ and magnon excitation frequencies $\omega_\text{m}$ are set by the perturbations that drive the magnetic order, and are therefore much smaller than the dominant energy scale $K$ responsible for stabilizing the Kitaev QSL. As such, dynamical signatures of incipient spin liquidity are generally expected in some range of $T$ and frequency $\omega$, sufficiently above $T_N$ and $\omega_\text{m}$, respectively.

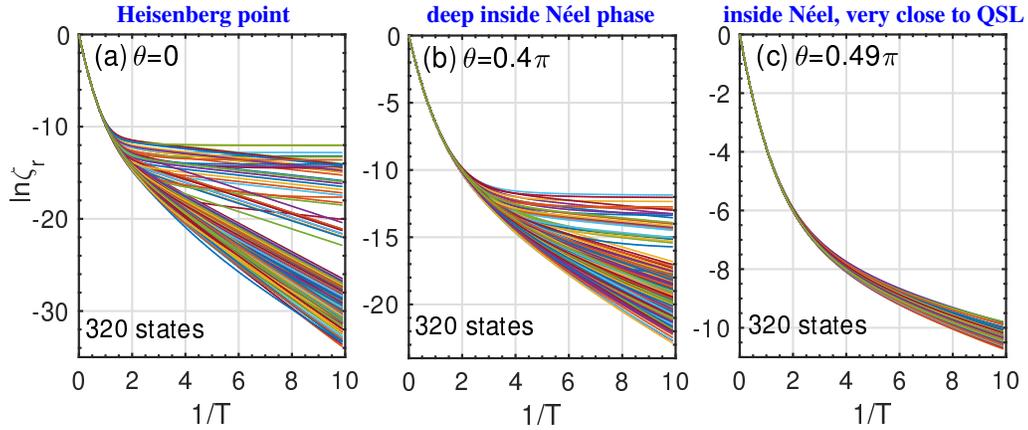
\begin{figure*}[!t] 
\vspace*{-6cm}
\centering\setlength{\unitlength}{\textwidth}
\begin{picture}(0.75,0.645)
\put(0,0){\includegraphics[width=0.75\textwidth,angle=0,clip=true,trim=0 0 0 0]{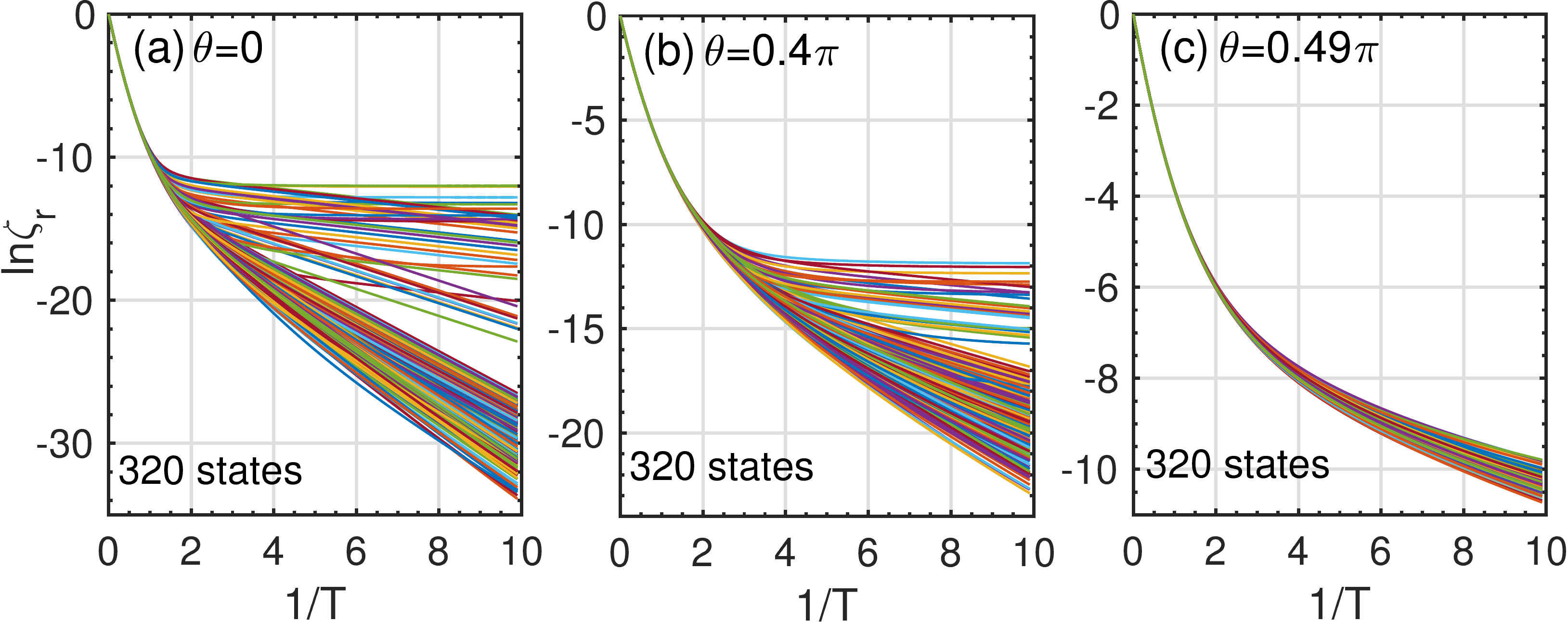}}
\put(0.08,0.3){{\bf {\cbl Heisenberg point}}}
\put(0.31,0.3){{\bf {\cbl deep inside N\'eel phase}}}
\put(0.53,0.3){{\bf {\cbl inside N\'eel, very close to QSL}}}
\end{picture}
\caption{{\bf Persistence of typicality near the Kitaev QSL, I.} Logarithm of contributions to the partition function $\zeta_r$ [Eq.~(\ref{eq:frt})] from 320 random states for three points inside the N\'eel state, stabilized by a Heisenberg coupling $J$: ({\bf a}) $\theta\!=\!0$, ({\bf b}) $\theta\!=\!0.4\pi$ and ({\bf c}) $\theta\!=\!0.49\pi$ [$J\!=\!\cos\theta$, $K\!=\!\sin\theta$]. 
All data refer to the symmetric 24-site cluster of Appendix~\ref{app:clusters}.}\label{fig:zetars}
\end{figure*}

As a dynamical probe we analyze the Raman scattering intensity, which measures the scattering of light off the magnetic degrees of freedom, as a function of energy and light polarization.~\cite{LF1968,Shastry1991,Devereaux2007} By monitoring the evolution of the intensity as we drive the honeycomb Kitaev QSL toward a number of different instabilities, we show that, irrespective of the nature of the phase the system is driven to, the response indeed follows closely the one originating from fractionalized quasiparticles in a wide range of $T$ and $\omega$, while the magnon-like response -- characteristic of the given magnetic order -- appears only at low $T$ and $\omega$ (see also \cite{Yamaji2018}). 
This crossover, which can be identified even relatively deep inside the ordered phase, is akin to the confinement of spinons in quasi-1D spin chains at low $T$.~\cite{Nagler91,Tennant93,Zheludev2000,Zaliznyak2004,Lake2005,Enderle2010,Mourigal2013,Bera2018}.

Besides the above dynamical crossover, our results reveal yet another manifestation of incipient spin liquidity, that of {\it persistent typicality}~\cite{Lloyd2013,Popescu2006,Goldstein2006,Reimann2007,Elsayed2013,Robin2014a,Robin2014b,Robin2016}: both static and dynamic properties can be captured down to surprisingly low $T$ by propagating, in real and imaginary time, a {\it single}, randomly chosen many-body quantum state. This remarkable property arises from the spectral weight downshift and the large, low-$T$ entropy characteristic of frustrated systems with a large number of competing low energy states and emergent gauge fields,~\cite{RamirezBook,Nasu2015,Yamaji2016} and has also been observed in the kagome Heisenberg antiferromagnet.~\cite{Sugiura2013}
The success of the typicality method is therefore inherently linked to the strong fluctuations present in quantum liquids. This therefore  opens a unique route to study, on a quantitative level, an abundance of strongly correlated materials, like, for example, the layered kagome antiferromagnet ZnCu$_3$(OH)$_6$Cl$_2$ or the honeycomb iridates and ruthenates, $\alpha$-Li$_2$IrO$_3$, Na$_2$IrO$_3$ and $\alpha$-RuCl$_3$.

\section{Model, Raman vertex and methods}\vspace*{-0.2cm}
We consider the generic situation of Fig.~\ref{fig:Cartoon} for the honeycomb magnet with Hamiltonian
\be
\mc{H} = K \sum\nolimits_{\alpha} \sum\nolimits_{\langle ij\rangle \in \alpha} S_i^\alpha S_j^\alpha + \mc{V}~.
\ee
Here, ${\bf S}_i$ and ${\bf S}_j$ are nearest-neighbor (NN) (pseudo-)spins-1/2, $K$ is the Kitaev coupling, $\alpha\!=\!x$, $y$, or $z$, depending on the orientation of the bond $\langle ij\rangle$ (see inset of Fig.~\ref{fig:Cartoon}), and $\mc{V}$ is a generic perturbation that destabilizes the Kitaev QSL. We will study here two such perturbations that are believed to be relevant in the available Kitaev materials, the NN Heisenberg exchange coupling, $J$, and the symmetric part of the NN off-diagonal exchange, $\Gamma$.~\cite{Katukuri2014,Rau2014,Satoshi2016,Sizyuk2014,Winter2016,Kee2016,Gamma2016} The corresponding models will be referred to below as the $JK$- and $K\Gamma$-model, respectively.

\begin{figure*}[!t] 
\includegraphics[width=0.75\textwidth,angle=0,clip=true,trim=0 0 0 0]{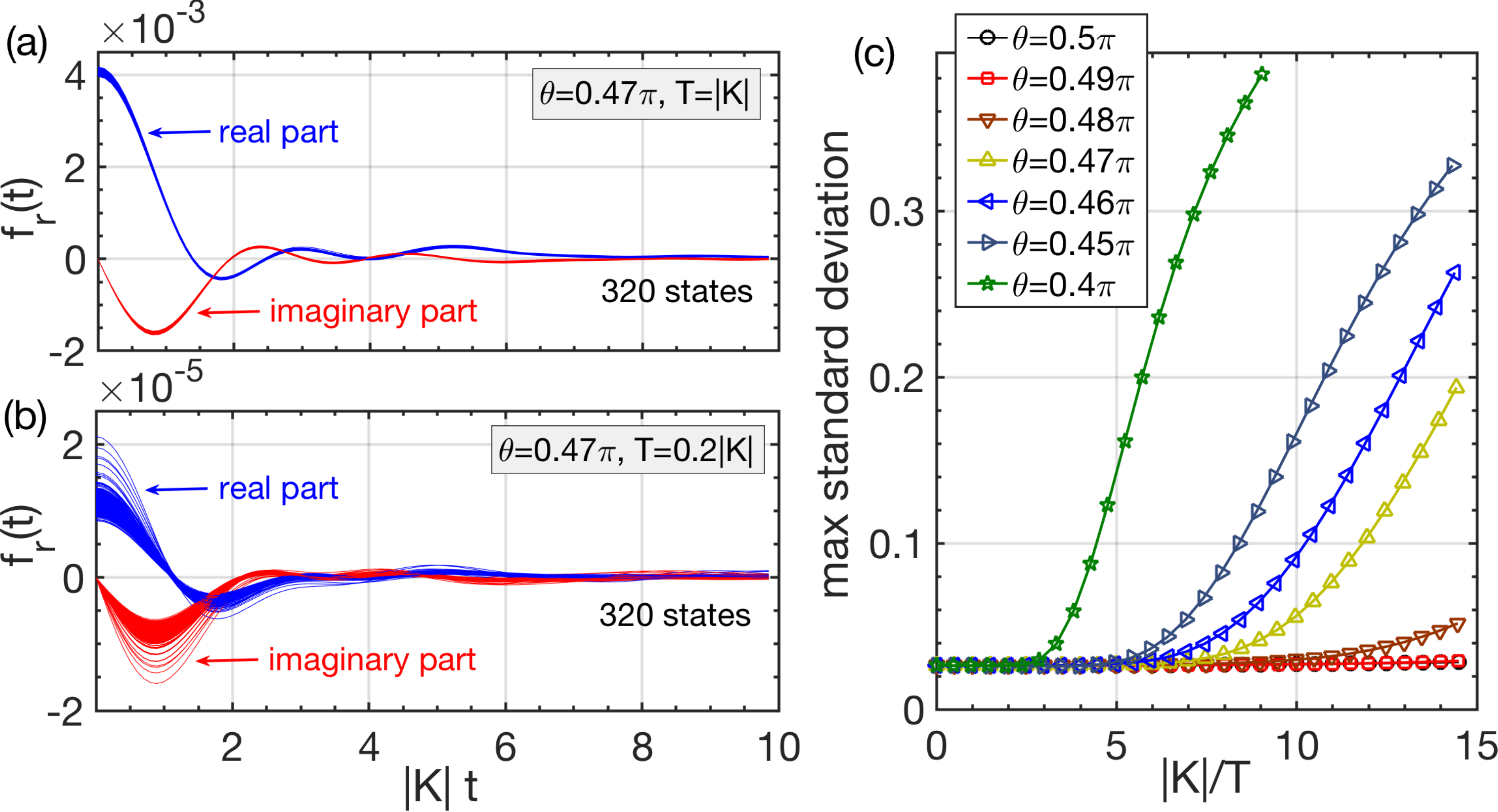}
\caption{
{\bf Persistence of typicality near the Kitaev QSL, II.} 
({\bf a}-{\bf b}) Real time evolution of the correlator $f_r(t)$ [Eq.~(\ref{eq:frt})] for 320 random ($T\!=\!\infty$) states $|r\rangle$, for the $JK$ model at $\theta\!=\!0.47\pi$, with $T\!=\!|K|$ ({\bf a}) and $0.2|K|$ ({\bf b}). 
({\bf c}) $T$-dependence of the maximum standard deviation in $\langle\mc{R}(t)\mc{R}(0)\rangle\!=\!\sum_r d_r f_r(t)/\sum_rd_r \zeta_r$ [see Eq.~(\ref{eq:frt})] for various values of $\theta$. 
All data refer to the symmetric 24-site cluster of Appendix~\ref{app:clusters}.
}
\label{fig:Error}
\end{figure*}

In the absence of $\mc{V}$, the Hamiltonian can be reduced to a quadratic problem of fermions moving in the background of static magnetic fluxes, and the ground state is a gapless QSL that lives in the zero-flux sector.~\cite{Kitaev2006} As shown by Nasu {\it et al},~\cite{Nasu2015} this state is adiabatically connected to the high-$T$ paramagnetic phase, but there are two characteristic crossover temperature scales, $T_\ell\!\sim\!K/100$ and $T_h\!\sim\!K$, see Fig.~2 of Ref.~[\onlinecite{Nasu2015}] and Fig.~\ref{fig:Cartoon}. Below $T_\ell$, the system is locked into the flux-free sector, while for $T_\ell\!<\!T\!<\!T_h$ the Majorana fermions move on top of a thermally disordered flux background.

Raman experiments measure the dynamical response function associated with the Raman vertex $\mc{R}$, which, in the Loudon-Fleury approximation,~\cite{LF1968,Shastry1991,Knolle2014} 
is given by
\be
\mc{R}=\sum\nolimits_{\langle ij\rangle} (\bs{\epsilon}_{\text{in}}\cdot{\bf d}_{ij})(\bs{\epsilon}_{\text{out}}\cdot{\bf d}_{ij})\mc{H}_{ij}\,,
\ee
where ${\bf d}_{ij}$ is the displacement from site $i$ to $j$,  $\bs{\epsilon}_{\text{in}}$ and $\bs{\epsilon}_{\text{out}}$ are the polarizations of the incoming and outgoing light, respectively, and $\mc{H}_{ij}$ denotes the interactions between $\vec{S}_i$ and $\vec{S}_j$. 
The Raman intensity is the Fourier transform $I(\omega)\!=\!\int\!d\omega  ~e^{i \omega t} \langle \mc{R}(t)\mc{R}_0\rangle$, where $\langle \cdots\rangle\!=\!\text{Tr}[e^{-\beta\mc{H}} \cdots ]/\text{Tr}[e^{-\beta\mc{H}}]$ denotes the statistical average over the Hilbert space, $\beta\!=\!1/T$ is the inverse temperature, and  $\mc{R}(t)\!=\!e^{i\mc{H}t} \mc{R}e^{-i\mc{H}t}$ is the time-evolved Raman vertex. 
In the 3-fold symmetric models we consider here, the Raman intensity comes entirely from the `xy' and `$x^2$-$y^2$' polarization channels. The two channels give identical contributions, as they belong to the two-dimensional irreducible representation $E_g$, so it suffices to consider the `xy' channel only.~\cite{Knolle2014,Brent2015}

The stochastic method used here amounts to replacing the thermodynamic trace over the Hilbert space of dimension $D$, with a sampling over $r_m$ randomly chosen states $|r\rangle$, with $r_m\!\ll\!D$.~\cite{Imada1986,Skilling1989,Drabold1993,Silver1994,Jaklic2000,Iitaka2003,Weisse2006,Prelovsek2013}
The Raman correlator is then given by
\be\label{eq:frt}
\begin{array}{c}
\langle\mc{R}(t)\mc{R}(0)\rangle\!\approx\!\sum_{r=1}^{r_m} d_r f_{r}(t)/\sum_{r=1}^{r_m} d_r \zeta_r, ~~\text{where}\\
\\
f_r(t)\!=\!\langle r| e^{-\beta \mc{H}/2}\mc{R}(t)\mc{R}(0)e^{-\beta \mc{H}/2}|r\rangle,~~~
\zeta_r\!=\!\langle r| e^{-\beta \mc{H}}|r\rangle~,
\end{array}
\ee
and $d_r$ is the dimensionality of the symmetry sector the given state $|r\rangle$ belongs to. 
The stochastic sampling is the basis of finite-$T$ Lanczos,~\cite{Jaklic2000,Prelovsek2013,Schnack2018} Chebyshev polynomial methods,~\cite{Weisse2006} and the standard typicality approach,~\cite{Popescu2006,Goldstein2006,Reimann2007,Elsayed2013,Robin2014a,Robin2014b,Robin2016} for which $r_m\!=\!1$.
Here we take $r_m\!=\!320$ random states (10 for each of the 32 irreducible representations of the symmetry group exploited for the symmetric 24-site cluster, see Appendix~\ref{app:clusters}). 
These states represent configurations at $T\!=\!\infty$, which then need to be propagated in real and imaginary time in order to evaluate their contribution, $f_r(t)$ and $\zeta_r$, to the dynamical correlation function and the partition function, respectively. This is done here using the standard Lanczos method.~\cite{Lanczos1950,Paige1971,CullumWilloughby1,Saad2011}  
The results are cross-checked with the complementary low-$T$ Lanczos method,~\cite{Aichhorn2003} which approaches the problem from low $T$, see Appendix~\ref{app:LTLM}. 


\section{Typicality}\vspace*{-0.2cm}
We begin by showing explicitly that the typicality hypothesis remains valid down to very low $T$ in the vicinity of the Kitaev QSL points. Fig.~\ref{fig:zetars} shows the logarithm of $\zeta_r$ defined in Eq.~(\ref{eq:frt}) for 320 states $|r\rangle$, for the $JK$-model at $\theta\!=\!0$ ({\bf a}), $0.4\pi$ ({\bf b}) and $0.49\pi$ ({\bf c}), where $J\!=\!\cos\theta$ and $K\!=\!\sin\theta$. All three panels correspond to points inside the N\'eel phase (the transition to the Kitaev QSL occurs at $\theta\!\simeq\!0.493\pi$, {see Ref.~[\onlinecite{Jackeli2013}] and Appendix~\ref{app:Boundaries}}). 
At the Heisenberg point ({\bf a}), the 320 random states give almost identical results for $\zeta_r$ for $T\!\gtrsim\!|K|$. This is the essence of the typicality hypothesis which tell us that states in the middle part of the energy spectrum are `typical' to each other, and therefore the statistical average can be equivalently obtained by looking at the evolution of a single random state. On general grounds, this hypothesis can be shown~\cite{Jaklic2000,DeRaedt2000,Prelovsek2013} to work extremely well at very high $T$. On cooling down, the system begins to sample the lower end of the energy spectrum, where finite-size effects begin to play a role and the eigenstates are usually in practise no longer typical. As a result, deviations between the different $\zeta_r$ become apparent. Quite remarkably, however, the characteristic temperature, $T_{\text{typ}}$, below which the typicality hypothesis breaks down gets lower and lower as we approach the Kitaev QSL point, see Fig.~\ref{fig:zetars}\,({\bf b}-{\bf c}) and sketch in Fig.~\ref{fig:Cartoon}.

This persistent typicality can also be seen in the time dependent quantities $f_r(t)$ defined in Eq.~(\ref{eq:frt}). Figs.~\ref{fig:Error}~({\bf a}-{\bf b}) show the real and imaginary parts of $f_r(t)$ for 320 states $|r\rangle$, for the representative point $\theta\!=\!0.47\pi$ of the $JK$-model. For $T\!=\!|K|$ ({\bf a}), the 320 random states give almost identical results for $f_r(t)$, in the entire time region shown. As above, deviations between the curves become visible at lower $T$, see panel {\bf b}. Nevertheless, as shown in panel {\bf c}, the maximum standard deviation between the different results for $\langle \mc{R}(t)\mc{R}(0)\rangle$ remains small down to very low $T$, in the vicinity of the Kitaev point.

\begin{figure*}[!t] 
\centering
\setlength{\unitlength}{\textwidth}
\begin{picture}(1,0.5)
\put(0,0){\includegraphics[width=0.95\textwidth,angle=0,clip=true,trim=0 0 0 0]{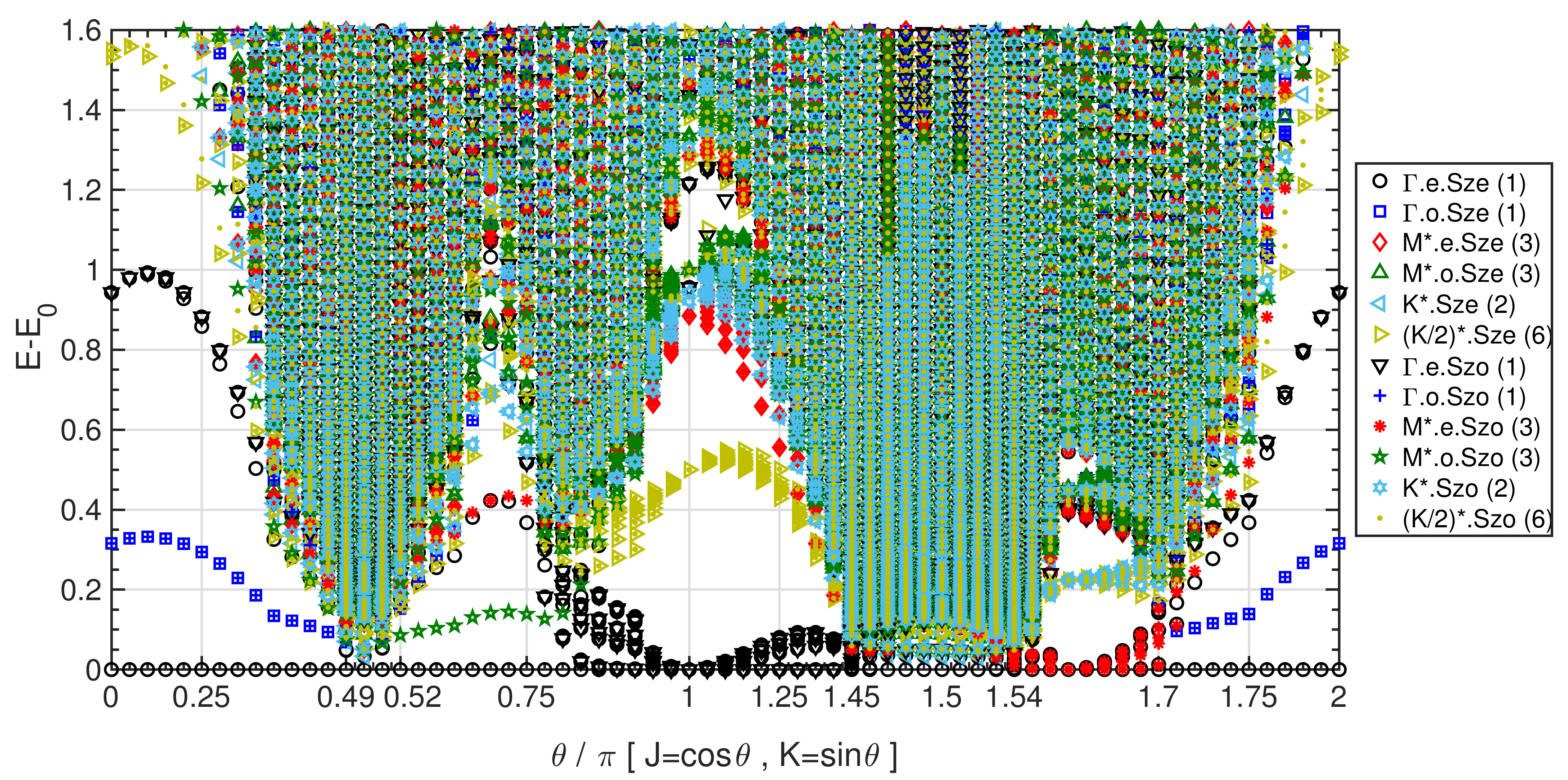}}
\put(0.067,0.46){\includegraphics[width=0.746\textwidth,angle=0,clip=true,trim=0 0 0 0]{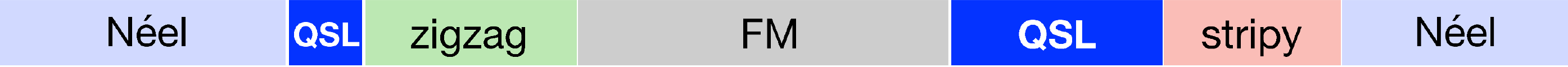}}
\put(0.83,0.38){\includegraphics[width=0.1\textwidth,angle=0,clip=true,trim=0 0 0 0]{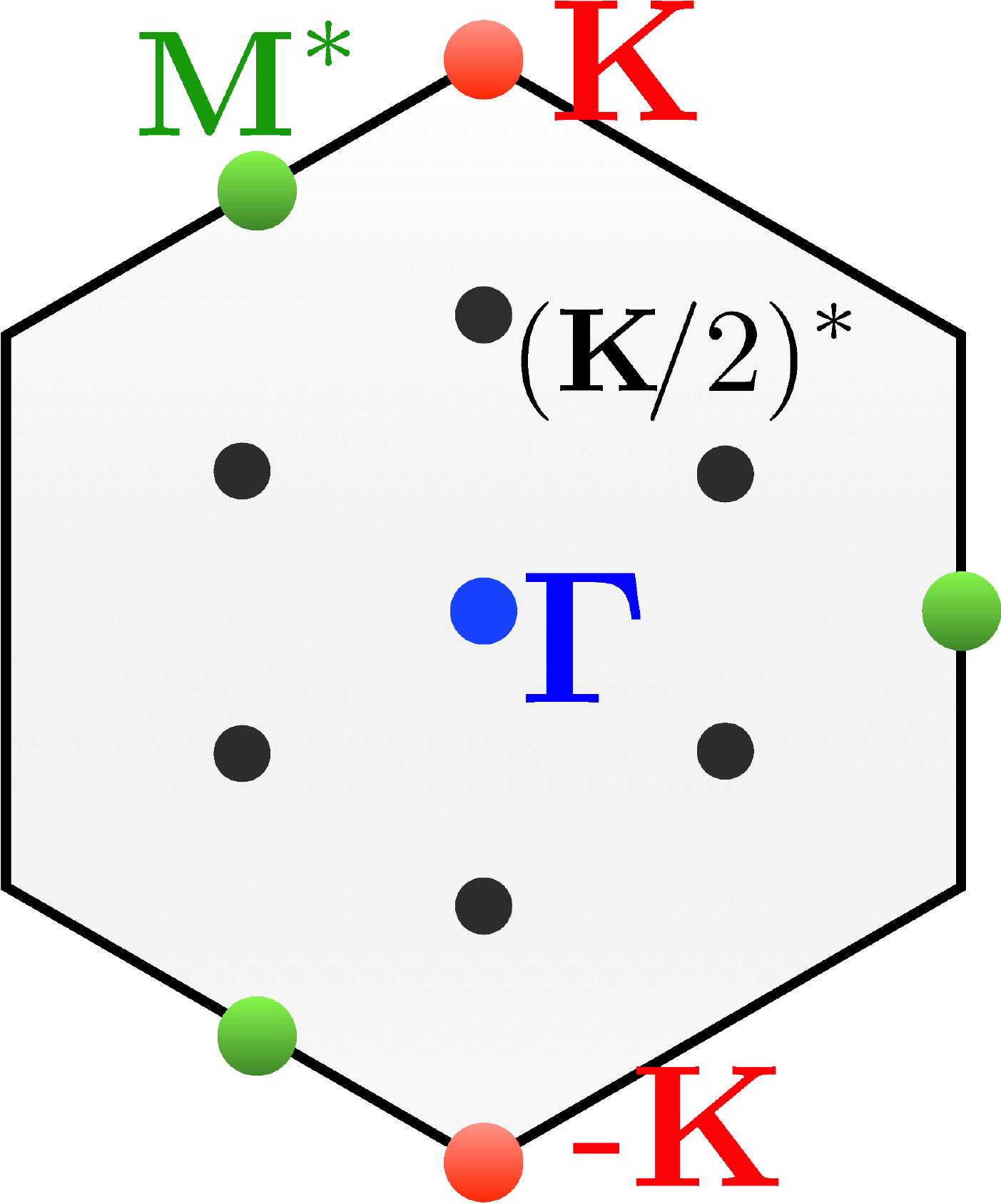}}
\end{picture}
\caption{{\bf Low-energy spectrum of the $JK$-model on the 24-site cluster.} Energies are measured from the ground state energy $E_0(\theta)$. The non-linear horizontal axis is used to better highlight what happens in the narrow regions around the Kitaev QSL points $\theta\!=\!\pi/2$ and $3\pi/2$. The various symbols correspond to the irreducible representations of the symmetry exploited here. The latter includes: i) the 12 translations, with the symbols $\Gamma$, $K^\ast$ and $M^\ast$ corresponding, respectively, to zero momentum, the corners of the first Brillouin zone (BZ), and the midpoints of the BZ edges. ii) real space inversion through the middle of the hexagons, with the letters `e' and `o' standing for the even and the odd sectors, respectively. iii) spin inversion (global rotation around the ${\bf x}$-axis in spin-space), with the symbols `Sze' and `Szo' standing for the even and odd sectors, respectively. The numbers in parentheses give the degeneracy of each level. The spectrum is obtained by the standard Lanczos method. Specifically, we show the 100 lowest energy states in each sector. Among these, only the lowest five are converged to the requested high precision (here relative precision in the 9th digit). The rest give a good representation with precision that lowers as we go up in energy.}\label{fig:Spectrum}
\end{figure*}

The success of the typicality method reflects, in essence, a fundamental property of the proximate Kitaev QSL, the presence of low-energy emergent gauge degrees of freedom. For conventional phases, the energy spectrum of a finite-size cluster is typically very dense (see below) in the middle of the spectrum but becomes sparse below some energy scale, proportionate to the bare interaction strength. For the Kitaev QSL, by contrast, the presence of magnetic flux sectors on one hand (with $2^{N/2}$ fermionic states each) and the very low flux gap ($\Delta\!\simeq\!0.065K$~\cite{Kitaev2006}) on the other, give rise to finite-size spectra that remain dense down to very low energies. This is demonstrated in Fig.~\ref{fig:Spectrum} which shows the energy spectrum of the $JK$-model on the 24-site cluster in the full parameter range of $\theta$. 
A direct consequence of the spectral downshift around the Kitaev points ($\theta\!=\!\pm\pi/2$) is that the system releases as much as half of its entropy only when cooled below $T\!\sim\!K/50$.~\cite{Nasu2015,Yamaji2016}
 
The exponential number of competing low-energy states in systems with emergent gauge fields quantifies what we mean by `very dense' spectrum. Indeed, as we demonstrate numerically in Appendix~\ref{app:typ}, the density of states $\rho(E)$ scales exponentially with system size $N$, 
\be\label{eq:rho}
\rho(E)\!\propto\!e^{N s(E)},
\ee 
where $s(E)$ is the microcanonical entropy per site (defined by $ds(E)/dE=\beta$), down to very low energies, already for clusters with 24 or 32 spins. 
This must be contrasted with the magnetically ordered regions of the phase diagram where, for the same finite-size systems, the spectral weight is concentrated near a restricted set of configurations (i.e., corresponding to elementary spin flips or magnons). 

The way in which the validity of relation (\ref{eq:rho}) for a finite-size system leads to persistent typicality is discussed at length and demonstrated explicitly based on numerical data in Appendix~\ref{app:typ}. A complementary way to understand the persistent typicality is to look at the mathematical predictions for the upper bound to the relative error $\delta B$ incurred in the statistical sampling of an operator $B$. Indeed, it can be shown~\cite{Jaklic2000,DeRaedt2000,Prelovsek2013,Sugiura2013} that $\delta B = \mc{O}(1/\sqrt{r_m \times Z_{\text{eff}}(T)})$, where $Z_{\text{eff}}=\text{Tr}[ e^{-\beta (\mc{H}-E_0)}]$ gives the effective number of thermally excited states at the given $T$, and $E_0$ is the ground state energy. 
It then follows that if (\ref{eq:rho}) holds for a given finite-size cluster and energy window (corresponding to $T$ via $ds(E)/dE=\beta$), then the relation $Z_{\text{eff}} \propto \rho(E)$ leads to an exponentially small upper bound in the error and, in turn, to persistent typicality. 
This observation has also been made by S.~Sugiura and A.~Shimizu in the context of the spin-1/2 kagome Heisenberg antiferromagnet.~\cite{Sugiura2013}


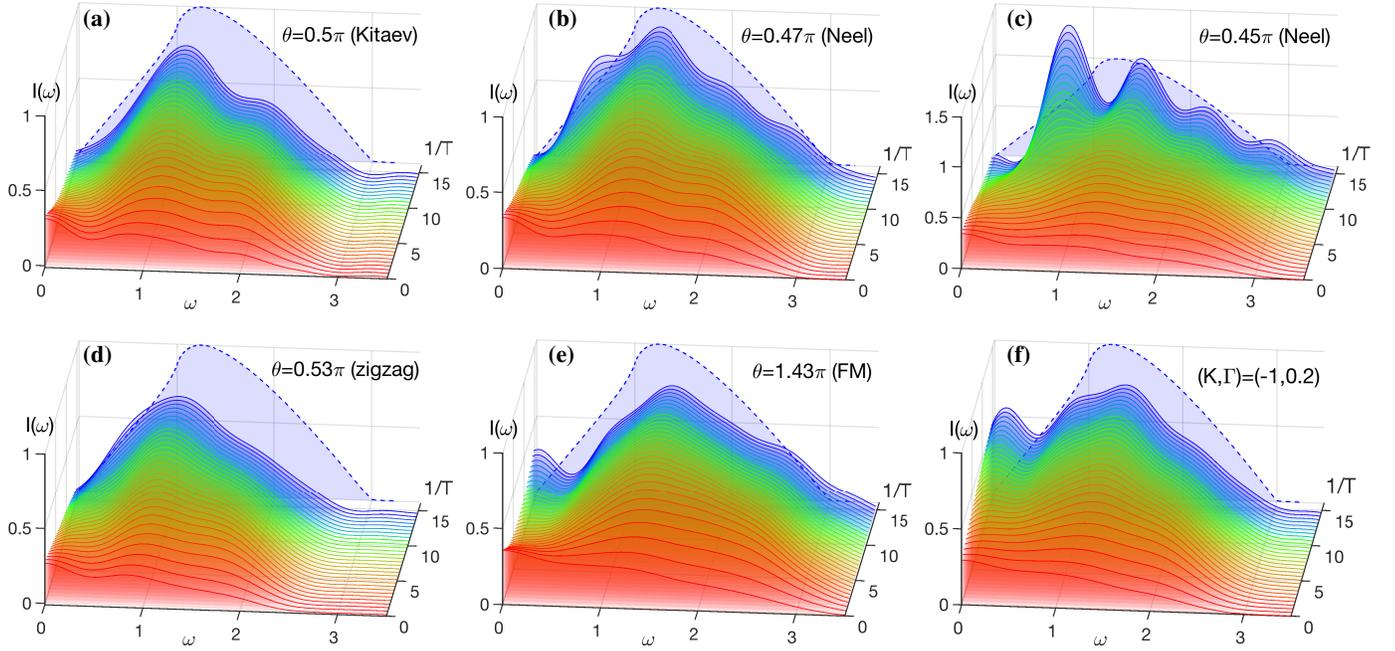
\begin{figure*}[!t] 
\vspace*{-2.75cm}
\centering
\setlength{\unitlength}{\textwidth}
\begin{picture}(1,0.645)
\put(0,0.25){
\includegraphics[width=0.33\textwidth,angle=0,clip=true,trim=0 0 0 0]{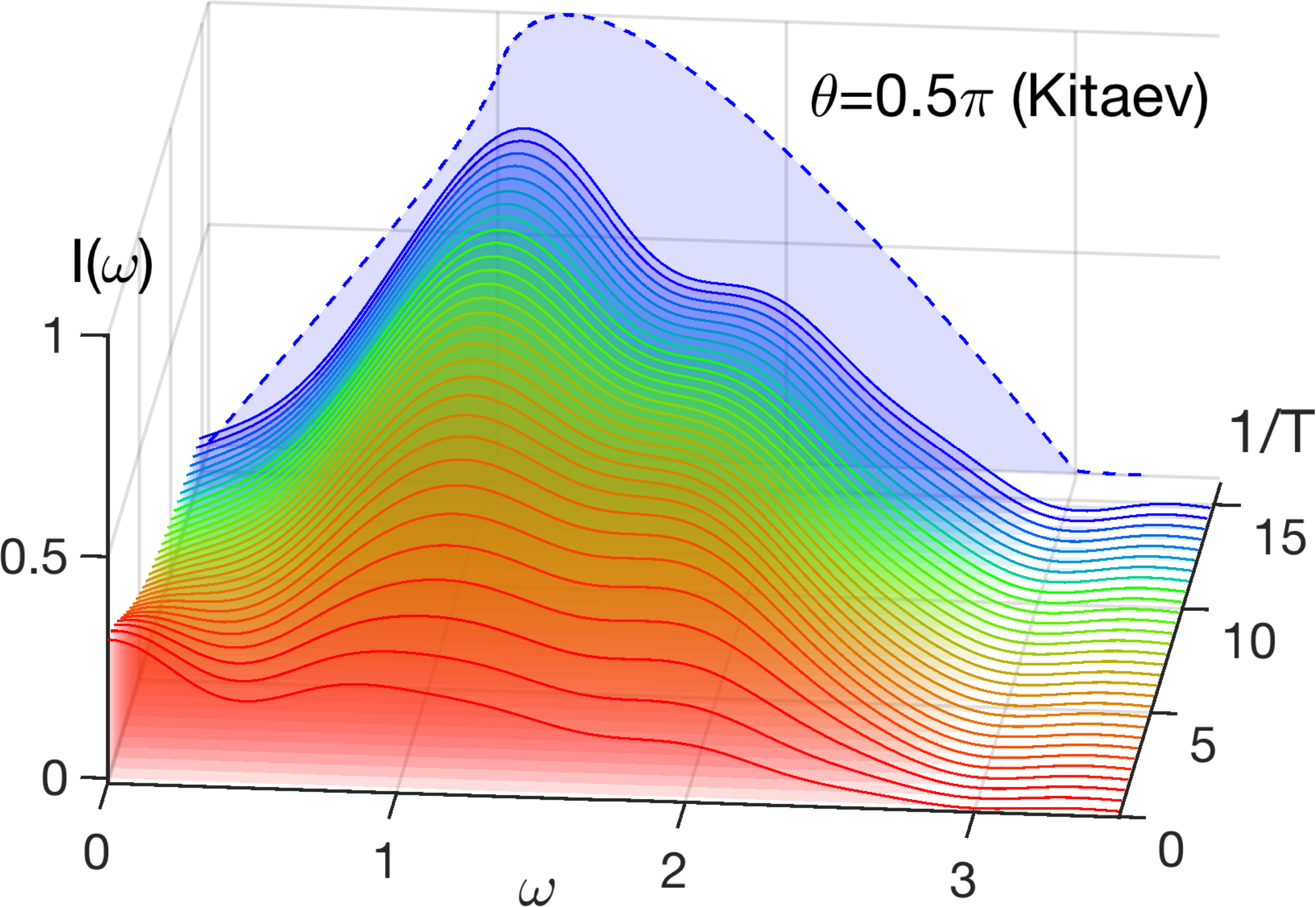}
}
\put(0.34,0.25){
\includegraphics[width=0.33\textwidth,angle=0,clip=true,trim=0 0 0 0]{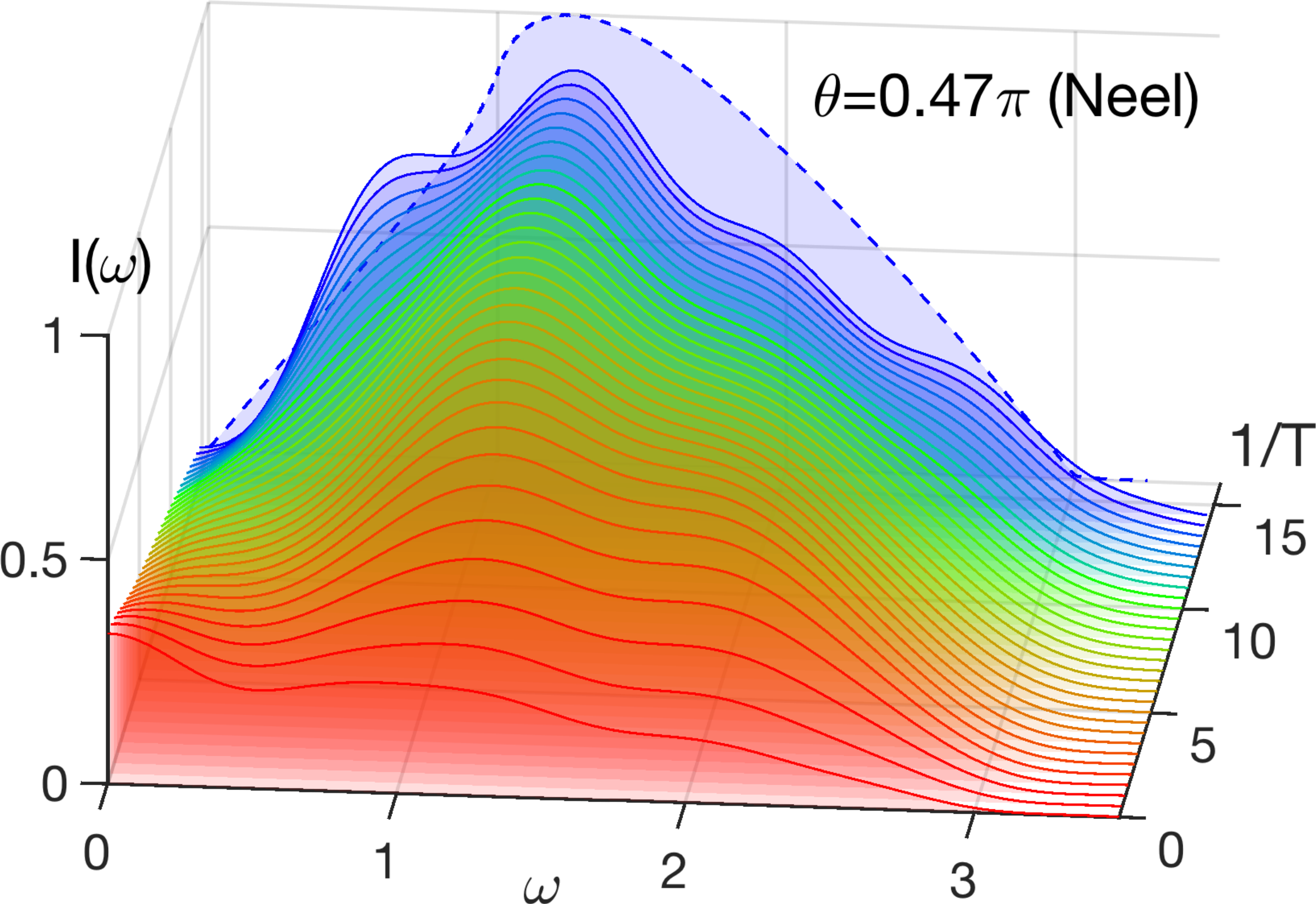}
}
\put(0.68,0.25){
\includegraphics[width=0.33\textwidth,angle=0,clip=true,trim=0 0 0 0]{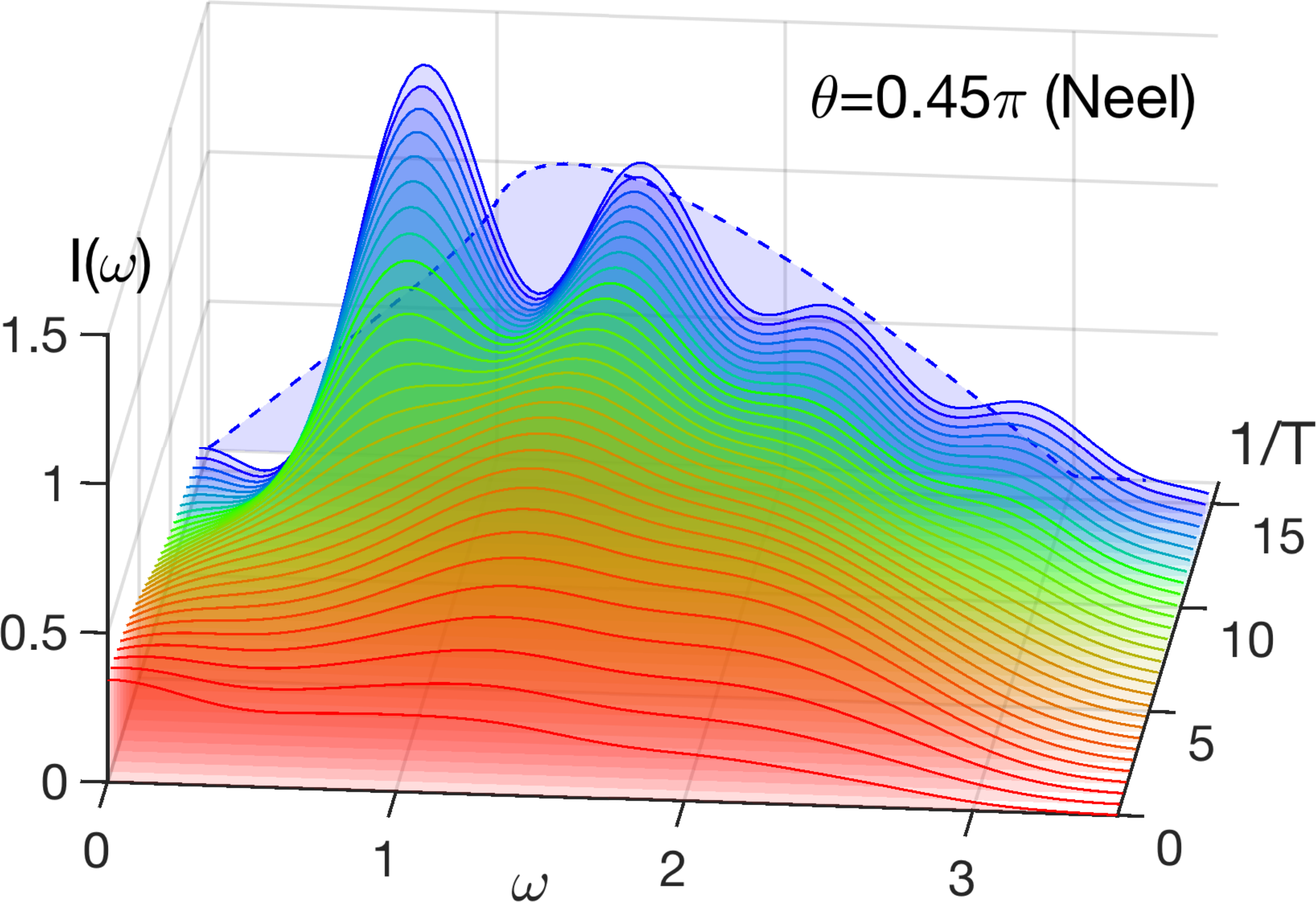}
}
\put(0,0){
\includegraphics[width=0.33\textwidth,angle=0,clip=true,trim=0 0 0 0]{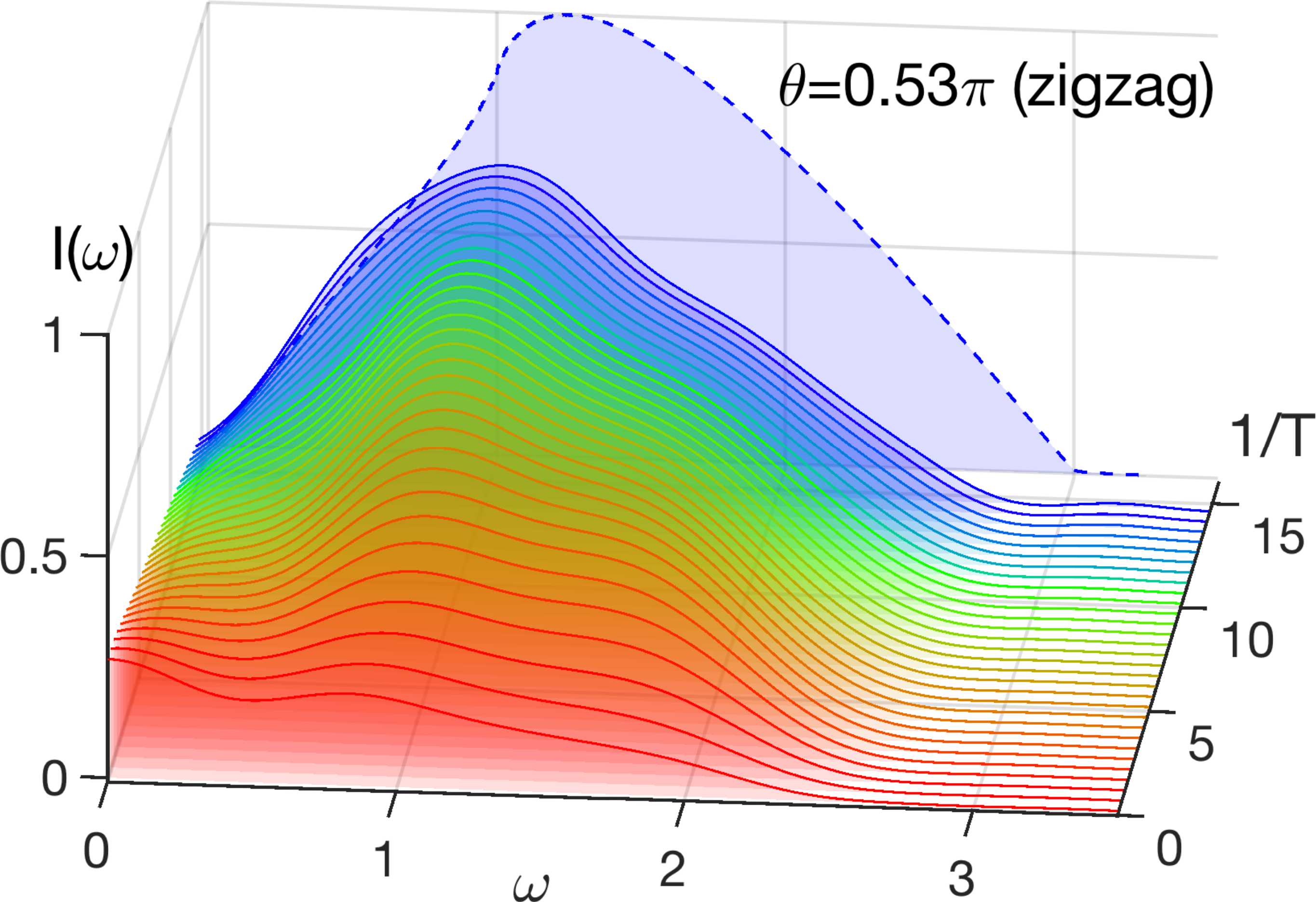}
}
\put(0.34,0){
\includegraphics[width=0.33\textwidth,angle=0,clip=true,trim=0 0 0 0]{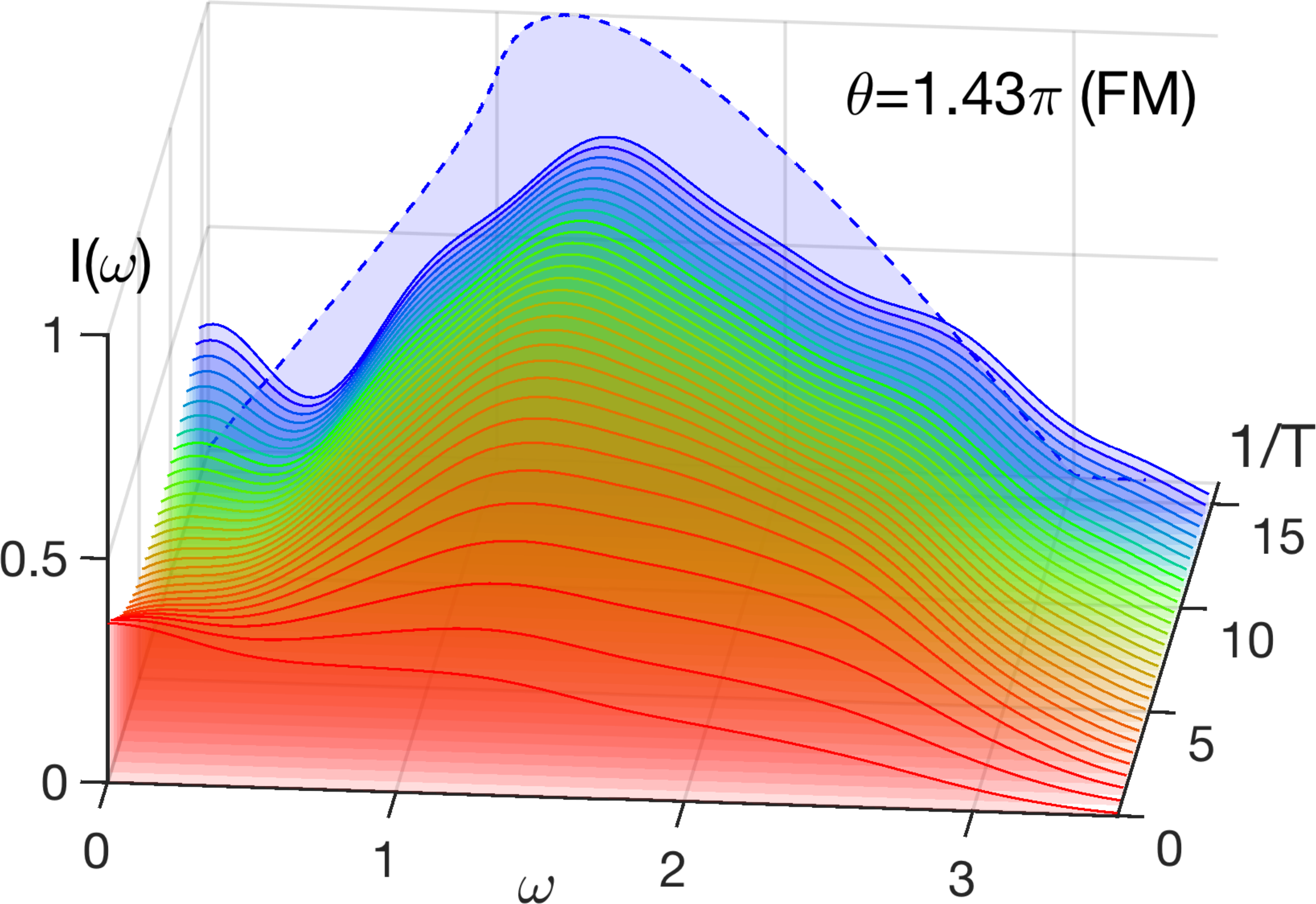}
}
\put(0.68,0){
\includegraphics[width=0.32\textwidth,angle=0,clip=true,trim=0 0 0 0]{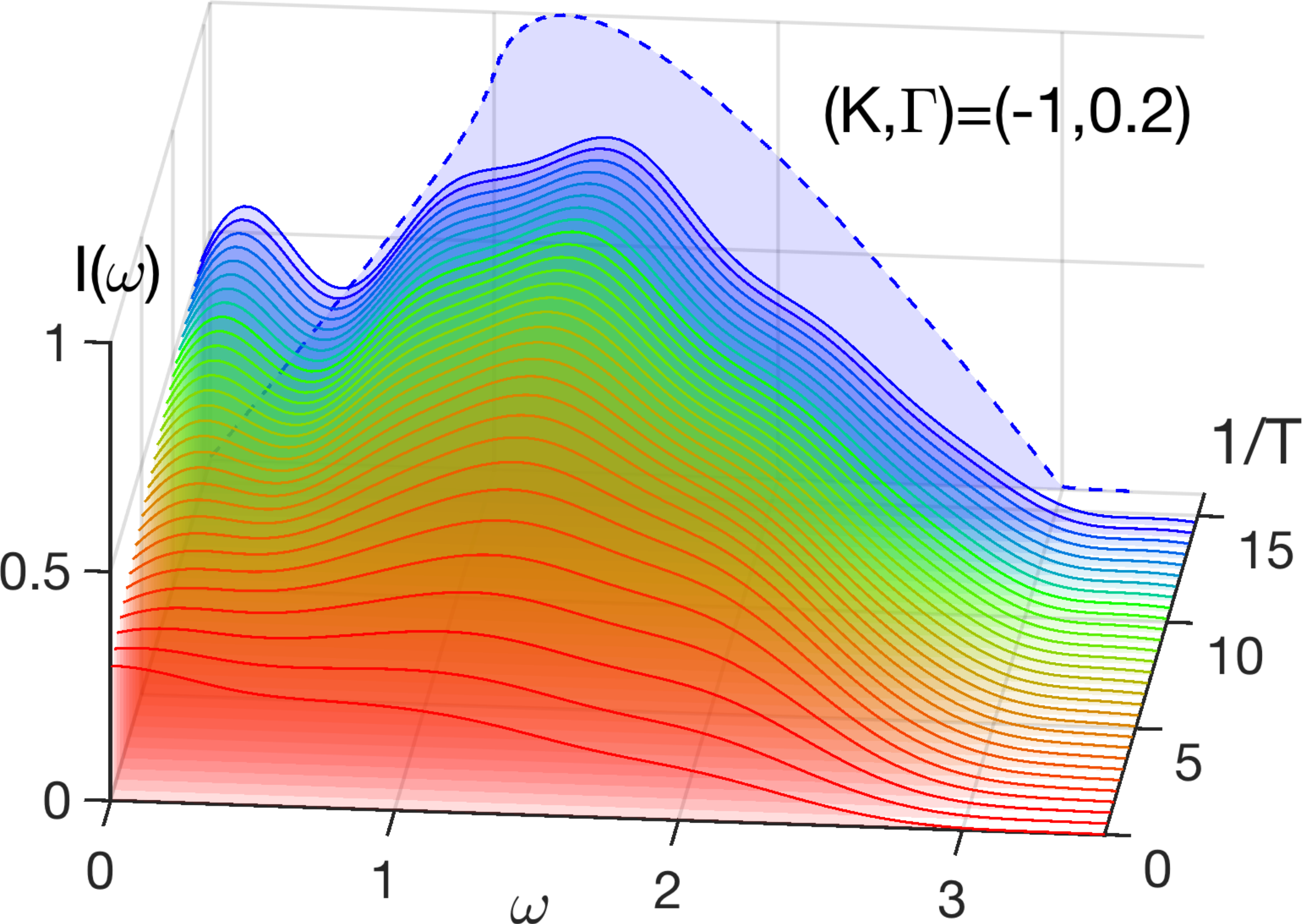}
}
\put(0.06,0.46){{\bf(a)}}
\put(0.405,0.46){{\bf(b)}}
\put(0.745,0.46){{\bf(c)}}
\put(0.06,0.21){{\bf(d)}}
\put(0.405,0.21){{\bf(e)}}
\put(0.745,0.21){{\bf(f)}}
\end{picture}
\caption{
{\bf Raman scattering intensity in the `xy' polarization channel for various phases proximate to the Kitaev QSL.} ({\bf a}) Ideal AF Kitaev point, ({\bf b}) $JK$ model with $\theta\!=\!0.47\pi$ (N\'eel phase), ({\bf c})  $JK$ model with $\theta\!=\!0.45\pi$ (N\'eel phase), ({\bf d})  $JK$ model with $\theta\!=\!0.53\pi$ (zigzag phase), ({\bf e}) $JK$ model with $\theta\!=\!1.43\pi$ (FM phase), ({\bf f}) $K\Gamma$ model with $K\!=\!-1$ and $\Gamma\!=\!0.2$.  
Dashed lines show the analytical $T\!=\!0$ result at the Kitaev QSL point.~\cite{Knolle2014}  
$T$ and $\omega$ are in units of $\sqrt{K^2\!+\!J^2}\!=\!1$ ({\bf a}-{\bf e}) or $|K|$ ({\bf f}). 
Results are obtained for the symmetric 24-site cluster of Appendix~\ref{app:clusters}, by propagating 320 random ($T\!=\!\infty$) states in real and imaginary time.
}\label{fig:Raman1}
\end{figure*}

\section{Dynamical crossover}\vspace*{-0.2cm}
We now turn to the dynamical crossover announced above. Fig.~\ref{fig:Raman1} shows the Raman intensities $I(\omega)$ for six different points in parameter space. Let us first focus on the first row (panels {\bf a}-{\bf c}), which show the evolution of $I(\omega)$ as we move deeper inside one particular magnetic phase, here the N\'eel state stabilized by a Heisenberg coupling $J$. As above, the $JK$-model is parametrized as $J\!=\!\cos\theta$ and $K\!=\!\sin\theta$, with $\theta\!=\!\pi/2$ (panel {\bf a}, QSL phase), $0.47\pi$ (panel {\bf b}, N\'eel) and $0.45\pi$ (panel {\bf c}, N\'eel). 
For the Kitaev point ({\bf a}), the results are consistent with both the exact $T\!=\!0$ result of Knolle {\it et al}~\cite{Knolle2014} (reference dashed lines) and the $T$-evolution data reported by Nasu {\it et al}.~\cite{Nasu2016} The broad profile with the main peak around $\omega\!\sim\!|K|$ (related to the Van Hove singularity of the two-fermion density of states) and the bandwidth of $3|K|$ are all reproduced, and the same is true for the characteristic growth of the zero-frequency intensity with $T$.~\cite{Nasu2016}

The crucial point is that the main features of the intensity remain the same in a wide $T$ and $\omega$ regime also inside the N\'eel phase. Additional structure, characteristic of this phase (and the given finite-size cluster), does become visible at low enough $T$, and becomes sharper as we move deeper inside the phase. 
In panel {\bf c}, for example, the main peak of the Kitaev point has been depleted, and the spectral weight has been asymmetrically transferred into two new peaks (one around $0.75|K|$ and the other around $1.5|K|$). At the same time, there is an appreciable gap developing at low $\omega$, while a fourth peak appears a little below $3|K|$. 
{While these sharper features may depend in detail on the finite-size cluster at hand,~\footnote{In a finite-size cluster, there are only a certain number of discrete momenta available. The low-$T$ Raman response arises then from pairs of magnon modes with total momentum zero, leading to a series of peaks characteristic of the finite-size cluster.} 
they are washed out with increasing $T$, and the response eventually resembles that at the Kitaev point (panel {\bf a}).}

Panels {\bf d}-{\bf f} show that this picture remains essentially the same irrespective of the nature of the phase proximate to the QSL. Here we show the intensity for three more phases, the zigzag ({\bf d}), the ferromagnetic ({\bf e}), and the phase stabilized by the off-diagonal exchange $\Gamma$ (for the nature of this phase see Refs.~[\onlinecite{Catuneanu2018,Gohlke2018,IoannisKGamma}]). 
As before, at sufficiently high $T$ the response is the same with that of the proximate Kitaev QSL, while the response at low $T$ is distinctive for each of the three phases. For example, the FM phase shows a relatively large zero-frequency weight, and the same is true for the phase stabilized by $\Gamma$.  
Results from the independent low-$T$ Lanczos method lead to the same conclusion, see Appendix~\ref{app:LTLM}.
Altogether, these results demonstrate the presence of a dynamical crossover from a conventional, magnon-like picture at low $T$ and $\omega$, characteristic of the low-$T$ ordered phase, to long-lived fractionalized quasi-particles, characteristic of the proximate Kitaev QSL. 


\section{Fermionic character}\vspace*{-0.2cm}
We now turn to the nature of these fractionalized quasiparticles and show evidence for their fermionic character. 
To this end, we follow the analysis of Ref.~\cite{Nasu2016} for the Raman intensity at the ideal QSL point. 
At this point, the Raman vertex is diagonal in the emergent fluxes {and only} excites/deexcites pairs of Majorana fermions. In particular, there are two processes contributing to the intensity, one (type A) corresponding to the creation of two fermions, and another (type B) to the creation of one fermion and the annihilation of another. The amplitudes of these processes then scale as $A\!\propto\![1-f(\varepsilon_1)][1-f(\varepsilon_2)]$ and $B\!\propto\! f(\varepsilon_1)[1-f(\varepsilon_2)]$, respectively, where $\epsilon_{1,2}$ are the energies of the two fermions involved in the process and $f(\varepsilon)\!=\!1/[1\!+\!\exp(\varepsilon/T)]$ is the Fermi-Dirac distribution.  
Furthermore, type (A) processes were shown to dominate the response at high frequencies, while type (B) dominate the response at low frequencies. 
Following the steps outlined in Ref.~\cite{Nasu2016} then, we consider the $T$-dependence of the integrated intensities, $n_L\!=\!\int_0^{\omega_1}\! d\omega ~\mc{I}(\omega)$ and $n_H\!=\!\int_{\omega_2}^{\infty}\! d\omega ~\mc{I}(\omega)$, with $\omega_1\!=\!0.25$ and $\omega_2\!=\!0.5$, and fit our numerical results for $n_L$ and $n_H$ to the expressions, respectively,  
\be\label{eq:yLH}
y_L \!=\! a_L f(\varepsilon_L^\ast) [1-f(\varepsilon_L^\ast)]+b_L,~~
y_H \!=\! a_H [1-f(\varepsilon_H^\ast)]^2+b_H,
\ee
where $a_L$, $b_L$, $a_H$, and $b_H$ are fitting parameters, and $\varepsilon_L^\ast\!=\!0.42$, $\varepsilon_H^\ast\!=\!0.58$ (see detailed justification for the choice of these parameters (and $\omega_{1,2}$ above) in Ref.~[\onlinecite{Nasu2016}]).

\begin{figure*}[!t] 
\centering
\includegraphics[width=0.95\textwidth,angle=0,clip=true,trim=0 0 0 0]{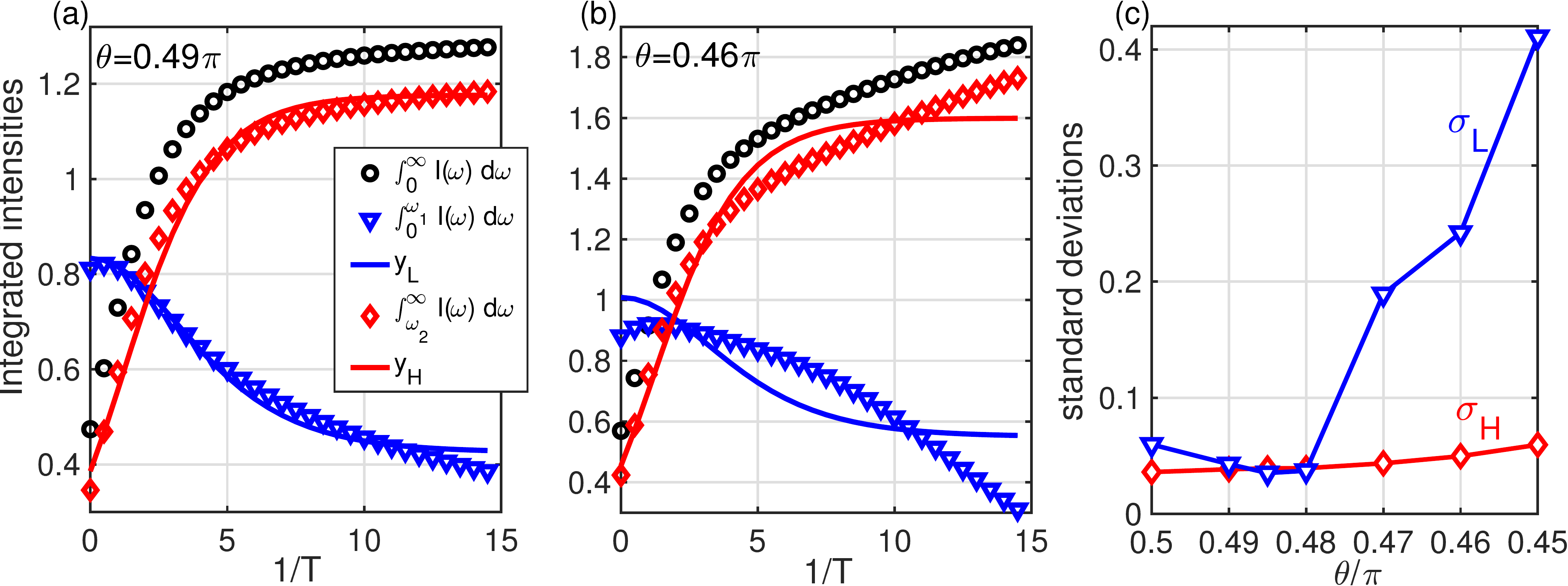}
\caption{
{\bf Fermionic character of the Raman response.} 
({\bf a}-{\bf b}) $T$-dependence of the integrated Raman intensities $n_{\text{tot}}\!=\!\int_0^\infty \!d\omega ~I(\omega)$ (black circles), $n_{L}\!=\!\int_0^{\omega_1} \!d\omega~ I(\omega)$ (blue triangles) and $n_{H}\!=\!\int_{\omega_2}^\infty \!d\omega ~I(\omega)$ (red diamonds), where $\omega_1\!=\!0.25$, $\omega_2\!=\!0.5$ (as in Ref.~[\onlinecite{Nasu2016}]), for the $JK$ model with $\theta\!=\!0.49\pi$ ({\bf a}) and $0.46\pi$ ({\bf b}). Solid lines are fits to Eqs.~(\ref{eq:yLH}), with $\varepsilon_L^\ast\!=\!0.42$ and $\varepsilon_H^\ast\!=\!0.58$ (as in Ref.~[\onlinecite{Nasu2016}]) with $a_L$, $b_L$, $a_H$ and $b_H$ determined by a least squared fitting procedure. 
({\bf c}) The standard deviations of the fits, $\sigma_L\!=\!\sqrt{\sum_{i=1}^p (n_{L}(T_i)\!-\!y_L(T_i))^2/p}$, and similarly for $\sigma_H$ [where $y_L$ and $y_H$ are defined in Eq.~(\ref{eq:yLH})], as we depart from the AF Kitaev point in the JK model parametrised as $K\!=\!\sin\theta$, $J\!=\!\cos\theta$.}\label{fig:Fits}
\end{figure*}

Figs.~\ref{fig:Fits} ({\bf a}-{\bf b}) show two representative fits of $n_L$ and $n_H$, obtained from a least-squared procedure, for two parameter points inside the N\'eel phase, one very close to the boundary with the Kitaev QSL phase ({\bf a}) and another further away ({\bf b}). The agreement between our numerical data for the integrated intensities and the approximate fermionic expressions of Eq.~(\ref{eq:yLH}) is very satisfactory close to the boundary, for both $n_L$ and $n_H$. As we depart further away from the boundary, however, the agreement remains good only for $n_H$, showing that the low-frequency part of the response is now controlled by magnon-like processes with different quantum statistical properties,  while the high-frequency part still tracks the behavior expected for fractionalized fermionic excitations. 

This picture is more vividly presented in panel {\bf c} which shows the evolution of the normalized standard deviations for the two fits, $\sigma_L$ and $\sigma_H$ (defined in the caption of Fig.~\ref{fig:Fits}), in the $JK$-model, as a function of $\theta$. 

\section{Discussion}\label{sec:Disc}
\vspace*{-0.2cm}
We have studied the Kitaev-Heisenberg honeycomb model, also with an additional $\Gamma$-term, using high- and low-temperature numerical approaches. We have found signatures of the fermionic Majorana quasi-particles, characteristic of the proximate Kitaev QSL, in the energy spectrum even when long-range magnetic ordering at very low $T$ is present. This is rather direct evidence that proximate quantum spin liquid physics can be observed at intermediate and high energies/frequencies even when the spectrum at low frequencies is strongly rearranged by a phase transition.

From a point of view of basic quantum statistical mechanics, it is worth commenting on the success of the typicality approach for describing the Kitaev quantum spin liquid. Even for system sizes as large as 24 spins, we had little trouble reaching reduced coupling strengths $K/T$ well in excess of 10. 
As we explain in more detail in  Appendix~\ref{app:typ}, to which we have relegated a proper technical discussion, the success of this method is based on two main ingredients. The first is the generation of a vector ($\frac{1}{\sqrt{\zeta_r}}e^{-\beta\mc{H}/2}|r\rangle$) corresponding to the coupling strength (or, in microcanonical terms, energy density) in question. This is achieved by subjecting a state $|r\rangle$, random in a local spin basis (Eq.~\ref{eq:randomstate}), to an imaginary time evolution (Eq.~\ref{eq:rbeta}).

The second ingredient is essentially equivalent with the eigenstate thermalisation hypothesis.~\cite{Srednicki1994,Deutsch1991,Rigol2008,Rigol2012,Rigol2016} This states that such a randomly chosen eigenvector at a given energy reproduces correlations characteristic of a thermodynamic ensemble at that energy.

Both of these ingredients are present if, for the finite-size cluster under consideration, the density of states remains sufficiently large at the energy in question. This is evidently a question of `detail', as the size of the eigenvectors that fit into the computer memory in practise correspond to sizes of a few dozen spins at best, a long way from the thermodynamic limit. 
The important general insight here is that highly frustrated systems -- by their very nature -- are ideal platforms for satisfying this condition (see also discussion in \cite{Sugiura2013}). In these systems, competing interactions lead to huge ground state degeneracies in simple idealised model systems.~\cite{MoessnerRamirez2006}

It is this huge ground state degeneracy which underpins much of the interest in frustrated magnets, as it renders them unstable to a host of different correlated magnetic phases. Crucially,  such instabilities only lift the degeneracy of the idealised model on the scale of the perturbation generating them. The entropy of this degenerate manifold gets spread over only this scale. Compared to a conventional magnet, this shows up in a strong downshift of spectral weight to energies below that of the leading term in the Hamiltonian. In fact, this distinguishes frustrated magnetism from `low-dimensional' routes to the suppression of magnetic ordering, and indeed, this been proposed as a practical diagnostic for this class of materials.~\cite{RamirezBook}

From the present perspective, this spectral weight downshift is precisely what is needed for the typicality method to work. This therefore accounts for both the good convergence of the method in the case of the pure Kitaev model; and for its robustness in the case of proximate spin liquidity, where it successfully accounts for the physics above the energy scale of the perturbation. Amusingly, the breakdown of the typicality method therefore coincides with that of the proximate spin liquid. 

This immediately implies that for the study of  the `correlated paramagnetic' behaviour of frustrated magnets, the typicality method should be useful much more generally -- what's needed are fluctuations over an large manifold of low-lying states, e.g.\ those that obey some kind of local constraint. Well-known examples are the low-lying singlets seen in the kagome antiferromagnet~\cite{Lecheminant1997,Waldtmann1998} and their interpretation in terms of nearest-neighbour valence bond states,~\cite{Moessner2001,ZengElser1995,Mambrini2000,Misguich2003,Arnaud2018} or the two-in/two-out states in pyrochlore spin ice.~\cite{Bramwell2001,Gingras2014,Fennell2014,BookFM}

Finally, adapting a less methodological viewpoint, it is perhaps also interesting to try to extend the phenomena observed here beyond highly frustrated magnets, by viewing these phenomena more generally from the perspective of confinement of fractional quasiparticles. For instance, this could be a  useful analogy for the confinement of spinons in quasi-1D spin chain materials, like SrCo$_2$V$_2$O$_8$,~\cite{Bera2018} at very low $T$. Given that the elementary magnon excitations associated with magnetic order carry integer quantum numbers, it is tempting to think of magnons as bound states of fractionalized quasi-particles, and the ordering as a condensation of such bound states. In such a scenario, the confining potential between the quasi-particles would be most effective below $T_N$ whereas, above $T_N$, deconfined quasi-particles could remain  evident in the spectrum.  This latter regime may also be accessible to the typicality method. 
The capacity of the typicality method to approach the critical coupling of such a transition, here or in the 2D case, potentially relevant to iridates and ruthenates like $\alpha$-RuCl$_3$, also remains to be explored.

We hope that our study will motivate further detailed studies  to a broader range of candidate quantum spin liquids, frustrated magnets and disordered quantum magnets, opening a window to their finite- but low-temperature properties that are otherwise inaccessible.

\noindent
\begin{center}{\bf Acknowledgements}\end{center}
We thank W. Brenig, K. Burch, M. Daghofer, A. Honecker and X. Zotos for fruitful discussions. IR acknowledges the hospitality of the Max Planck Institute for the Physics of Complex Systems (MPI-PKS) of Dresden, where part of this work was done. This work was in part supported by the Deutsche Forschungsgemeinschaft via Grant No. SFB 1143. SK was partially supported through the Boston University Center for Non-Equilibrium Systems and Computation. IR and NBP were supported by the US Department of Energy, Office of Science, Basic Energy Sciences under Award No. DE-SC0018056.

\appendix
\titleformat{\section}{\normalfont\bfseries\filcenter}{Appendix~\thesection:}{0.25em}{}

\section{Typicality method}\label{app:typ}
\vspace*{-0.2cm}
As we discussed in the main text, one way to understand the mechanism behind the success of the typicality method is to look at the mathematical predictions for the upper bound of the relative error incurred in the stochastic sampling of an operator.~\cite{Jaklic2000,DeRaedt2000,Prelovsek2013,Sugiura2013} When combined with the characteristic spectral downshift and the exponentially large density of states down to low energies, these mathematical bounds can become exponentially small, leading to typicality. 
This connection has been made previously by S. Sugiura and A. Shimizu, in the context of the kagome Heisenberg antiferromagnet.~\cite{Sugiura2013}
Here, we shall attempt to shed some more light into this mechanism and, at the same time, demonstrate explicitly how and when different starting random vectors $|r\rangle$ can deliver indistinguishable results down to very low $T$. In particular, we shall demonstrate numerically the relation (\ref{eq:rho}), which is the basis for the success of the typicality method. We shall also discuss the connection to the eigenstate thermalization hypothesis (ETH).~\cite{Srednicki1994,Deutsch1991,Rigol2008,Rigol2012,Rigol2016} 

Let us start with a brief description of the method.  
For a given finite-size cluster (see next section), we exploit symmetries to reduce the size of the Hilbert space. Then, for each given symmetry sector, we choose random initial states  
\be
|r\rangle = \sum_{\ell=1}^D d_\ell~ |\ell\rangle / \sqrt{\xi},
\label{eq:randomstate}
\ee
where $D$ is the dimension of the Hilbert space inside the given sector, $\{|\ell\rangle, \ell=1\text{-}D\}$ is a basis of product-like states, and $\xi\!\!=\!\!D/3$ is a normalization constant. The coefficients $d_\ell$ are drawn with a uniform distribution in the unit circle, with
\be\label{eq:dldl}
\overline{d_\ell^\ast d_{\ell'}}=\overline{|d|^2} \delta_{\ell\ell'},~~\overline{|d|^2}=1/3.
\ee  
The analogous relation for the correlations between the coefficients $r_i\!=\!\langle i|r\rangle$, where $|i\rangle$ are the eigenstates of $\mc{H}$ with $\mc{H} |i\rangle \!=\! E_i |i\rangle$, is
\bea\label{eq:riri}
&&\overline{r_i^\ast r_{i'}}=\frac{1}{\xi}\sum_{\ell\ell'} \overline{d_\ell^\ast d_{\ell'}} ~\langle  \ell|i\rangle\langle  i'|\ell'\rangle = 1/D ~\delta_{ii'}.
\eea
We then propagate the random state $|r\rangle$ in imaginary time using the standard Lanczos method,~\cite{Lanczos1950,Paige1971,CullumWilloughby1,Saad2011} to obtain the state
\be
|r,\beta\rangle \equiv e^{-\beta\mc{H}/2} |r\rangle.
\label{eq:rbeta}
\ee
Roughly speaking, during the imaginary time evolution, {the coefficients $r_i$ are amplified (suppressed) depending on whether the corresponding} energy $E_i$ is below (above) $T\!=\!1/\beta$, respectively. Ideally, at large enough $\beta$, the normalized state $|r,\beta\rangle/\sqrt{\zeta_r}$  converges to the ground state of $\mc{H}$ within the given sector. The speed of convergence to the ground state is related to the sparseness of the spectrum, and the size of the first excitation gap in particular.~\cite{Saad2011} 

\begin{figure}[!t] 
\centering
\includegraphics[width=0.455\textwidth,angle=0,clip=true,trim=0 0 0 0]{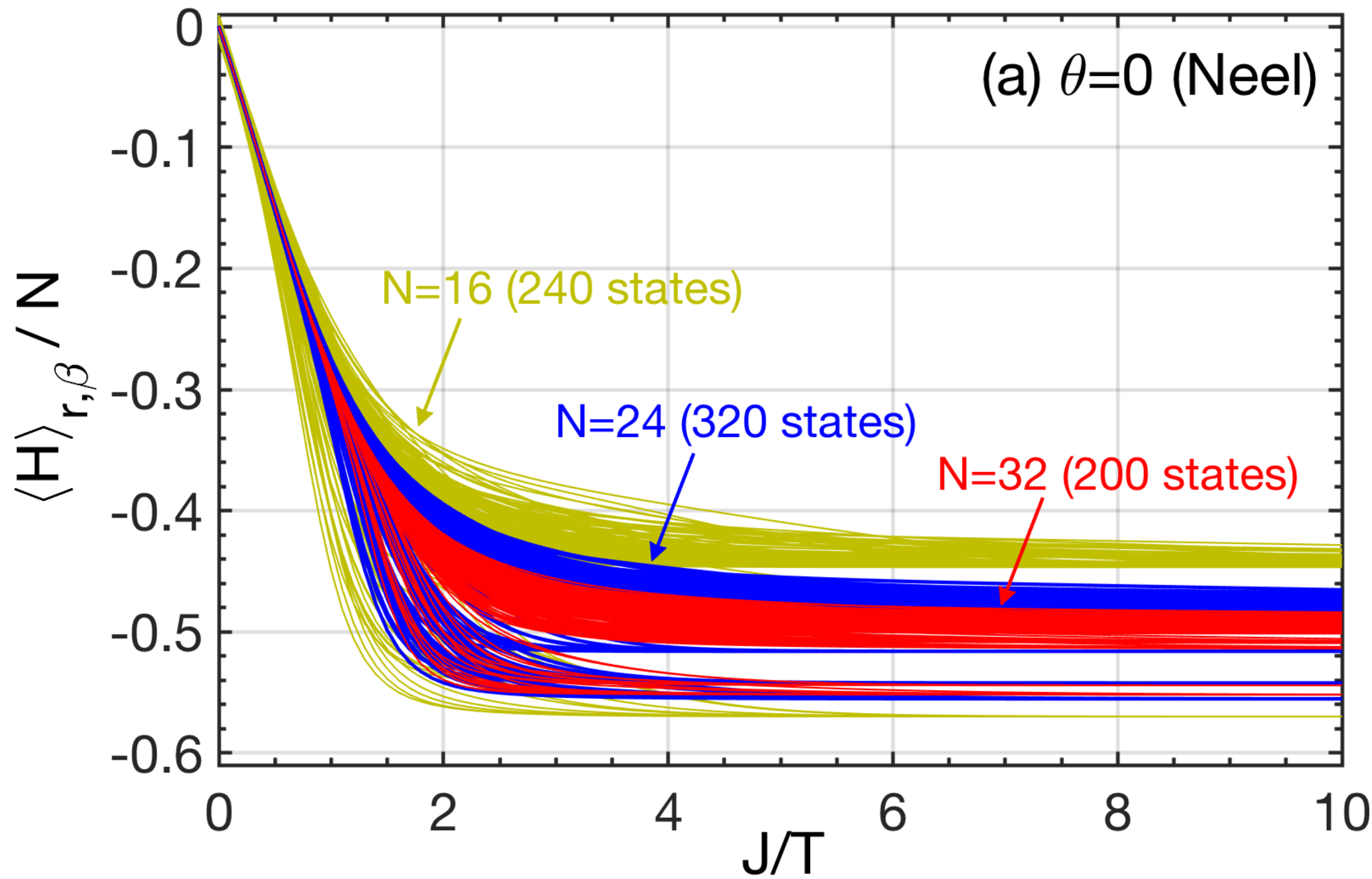}
\includegraphics[width=0.45\textwidth,angle=0,clip=true,trim=0 0 0 0]{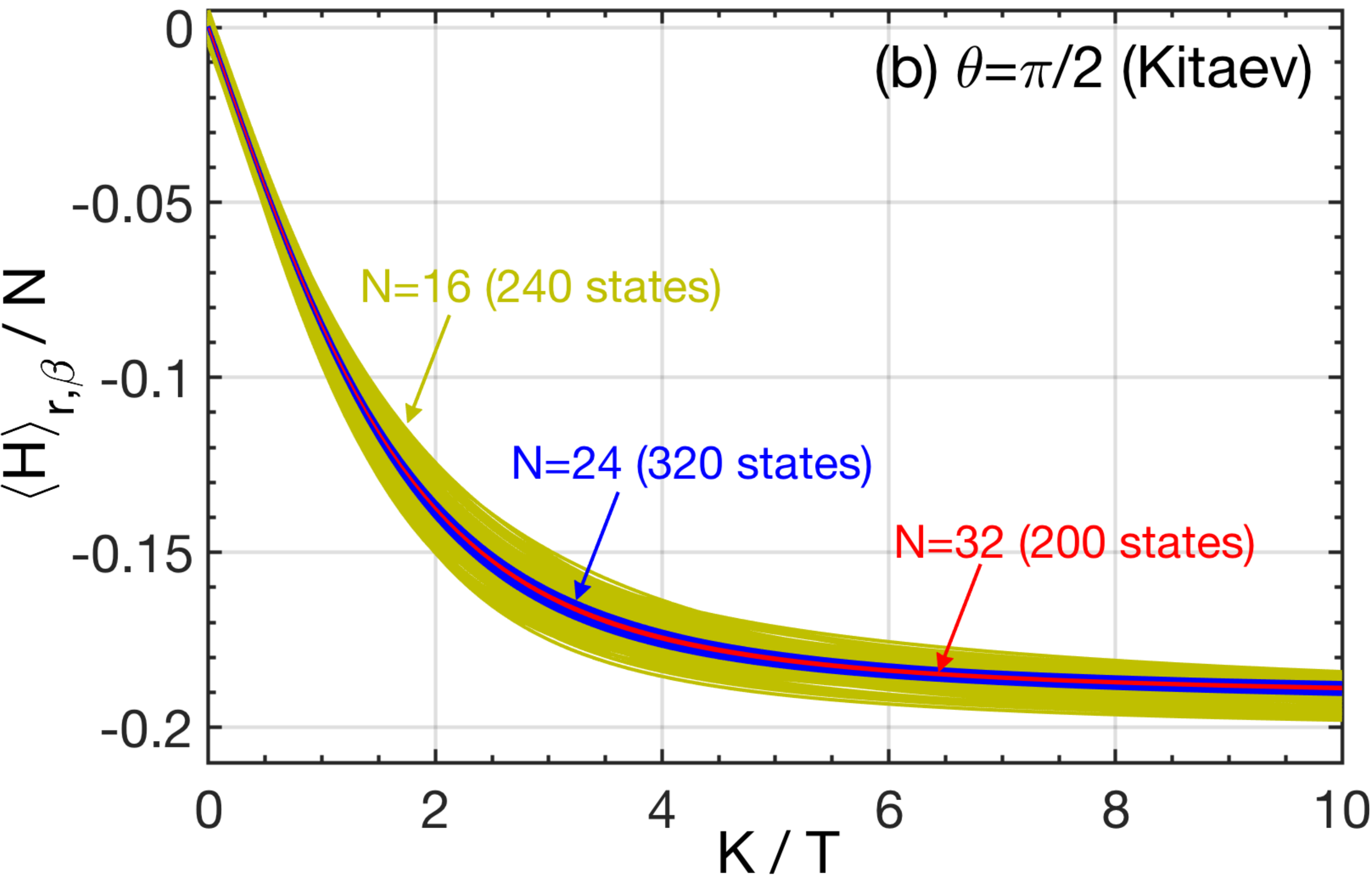}
\caption{
Expectation value of the energy per site, $\langle\mc{H}\rangle_{r,\beta}/N$, in the normalized state $|r,\beta\rangle/\sqrt{\zeta_r}$ [see definition in Eq.~(\ref{eq:EoN})] for a number of initial random states $|r\rangle$ (10 in each irreducible representation; In total, 240 states for the 16-site cluster, 320 for the 24-site cluster, and 200 for the 32-site cluster; for the latter we only show data from the sectors that are even under spin inversion).  Panels (a) and (b) correspond to the AF Heisenberg and the AF Kitaev point, respectively.
}\label{fig:EoN}
\end{figure}

Fig.~\ref{fig:EoN} shows the expectation value of $\mc{H}$ in $|r,\beta\rangle/\sqrt{\zeta_r}$,
\be\label{eq:EoN}
\langle \mc{H}\rangle_{r,\beta}\equiv
\langle r,\beta|\mc{H}|r,\beta\rangle/\zeta_r~,
\ee 
divided by system size $N$, for a large number of random states $|r\rangle$, for three system sizes ($N\!=\!16$, $24$ and $32$ sites), and for two special points in parameter space, the ideal AF Heisenberg point (panel {\bf a}) and the ideal AF Kitaev point (panel {\bf b}). 
For the N\'eel state [Fig.~\ref{fig:EoN}\,{\bf a}], the curves converge very quickly with increasing $\beta$, and the converged values correspond to the ground state energy for each given symmetry sector. In particular, the large spreading in the converged values for each given cluster arises: i) from the fact that we have taken random states in many different sectors (24, 32 and 20 sectors for the 16-, 24- and 32-site cluster, respectively) and, more importantly, ii) from {sparseness} of the energy spectrum at the N\'eel point, see Fig.~\ref{fig:Spectrum} of the main text. 

The situation is very different for the Kitaev point [Fig.~\ref{fig:EoN}\,{\bf b}]. Here, the convergence to the ground state is much slower compared to the N\'eel state, which reflects the very dense spectrum at the Kitaev point and, in particular, the tiny energy gap to the first excited state (see Fig.~\ref{fig:Spectrum}). Furthermore, the curves from all different initial states and all different clusters fall on top of each other, and the width of the distribution becomes narrower as we increase the system size. 

Let us now discuss why typicality works so well in the vicinity of the Kitaev points. 
There are two main ingredients behind the success of the typicality method. The first is that the state $|r,\beta\rangle$ is dominated by eigenstates $|i\rangle$, with $E_i$ lying inside a small window around the canonical expectation value $\mc{H}$, and in particular, that 
\be\label{eq:i}
\langle \mc{H}\rangle_{r,\beta} \approx \text{Tr}[e^{-\beta\mc{H}}\mc{H}]/\text{Tr}[e^{-\beta\mc{H}}]~.
\ee
The second ingredient for the success of typicality is that the expectation value of some local operators $A$ (static or dynamic) in the state $|r,\beta\rangle$ is the same with the result we get with the canonical ensemble, 
\be\label{eq:ii}
\langle A\rangle_{r,\beta}\equiv
\langle r,\beta|A |r,\beta\rangle/\zeta_r
\approx 
\text{Tr}[A e^{-\beta\mc{H}}]/\text{Tr}[e^{-\beta\mc{H}}]~.
\ee
Given the first ingredient, Eq.~(\ref{eq:ii}) is equivalent with the so-called eigenstate thermalization hypothesis (ETH).~\cite{Srednicki1994,Deutsch1991,Rigol2008,Rigol2012,Rigol2016} 
We will now show how both these ingredients are guaranteed if the density of states is sufficiently large. To this end, we rewrite 
\bea\label{eq:avgE1}
\zeta_r = \!\sum\nolimits_i |r_i|^2 e^{-\beta E_i},~~~
\langle \mc{H} \rangle_{r,\beta}=\!\sum\nolimits_i |r_i|^2 e^{-\beta E_i}E_i / \zeta_r~,
\eea
and then break up the sum over $i$ into a sum over energy windows $[E-\delta E/2, E+\delta E/2]$ around $E$ and a sum over states $j$ in each window, 
\be\label{eq:sum}
\begin{array}{c}
\zeta_r \to \sum_E \delta\mc{N}(E) ~e^{-\beta E} g_r(E),\\
\\
\langle \mc{H} \rangle_{r,\beta} \to 
\sum_E \delta\mc{N}(E) ~E ~e^{-\beta E} g_r(E) / \zeta_r,
\end{array}
\ee
where $\delta\mc{N}(E)=\rho(E)\delta E$ and the quantity 
\be\label{eq:grE}
g_r(E)=\sum_{j=1}^{\delta\mc{N}(E)}|r_{E,j}|^2/\delta\mc{N}(E),
\ee
is the average of $|r_i|^2$ inside the window around $E$.
The above way of rewriting the sums in (\ref{eq:avgE1}) as sums over energy windows has a meaning only if most of the coefficients $|r_{E,j}|^2$ are finite. 
If, in addition, $\delta\mc{N}(E)$ is large enough then we can use the central limit theorem to make a statement for the distribution of $g_r(E)$ (for fixed $E$).  
According to that theorem, in the limit of large $\delta\mc{N}(E)$, the distribution of $g_r(E)$ (for fixed $E$) approaches a normal distribution with standard deviation $\sigma_E^2/\delta\mc{N}(E)$, where $\sigma_E^2$ is the variance of the individual coefficients $|r_{E,j}|^2$. Therefore, if $\delta\mc{N}(E)$ is large we can replace
\be
|r_{E,j}|^2 \to \overline{|r|^2} = 1/D
\ee
in (\ref{eq:grE}), which in turn gives $g_r(E)\!\approx\!1/D$, i.e. $g_r(E)$ becomes independent of the random state $|r\rangle$, see below. 
The equivalence with the canonical ensemble follows immediately. 

We have already emphasised the need for a large enough density of states. In practice, we only need this condition to hold in a narrow region of energy that is fixed by $\beta$. Indeed, for large enough system sizes (and finite energy density $E/N$), the density of states $\rho(E)$ scales exponentially with  system size $N$, namely, $\rho(E)\!\propto\!e^{N s(E)}$, where $s(E)$ is the microcanonical entropy per site. Then, the sum over $E$ in (\ref{eq:sum}) is dominated by a very narrow region around a characteristic energy $E^\ast$ fixed by  $s'(E^\ast)\!=\!\beta$. 
Using the steepest descent method then gives, for example, 
\be\label{eq:zetar}
\zeta_r(\beta) \propto  \rho(E^\ast) e^{-\beta E^\ast}. 
\ee

\begin{figure}[!t] 
\centering
\includegraphics[width=0.45\textwidth,angle=0,clip=true,trim=0 0 0 0]{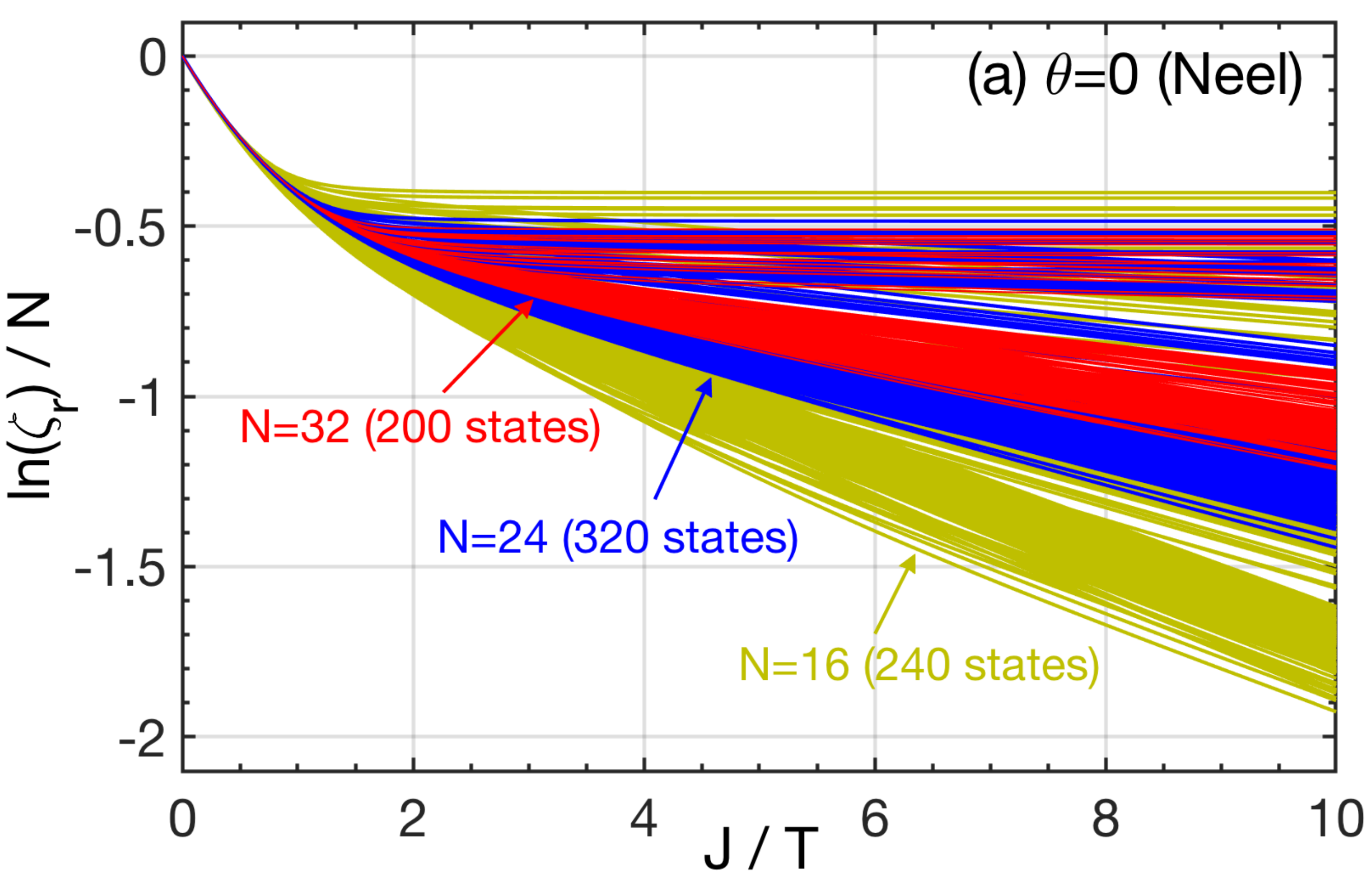}
\includegraphics[width=0.45\textwidth,angle=0,clip=true,trim=0 0 0 0]{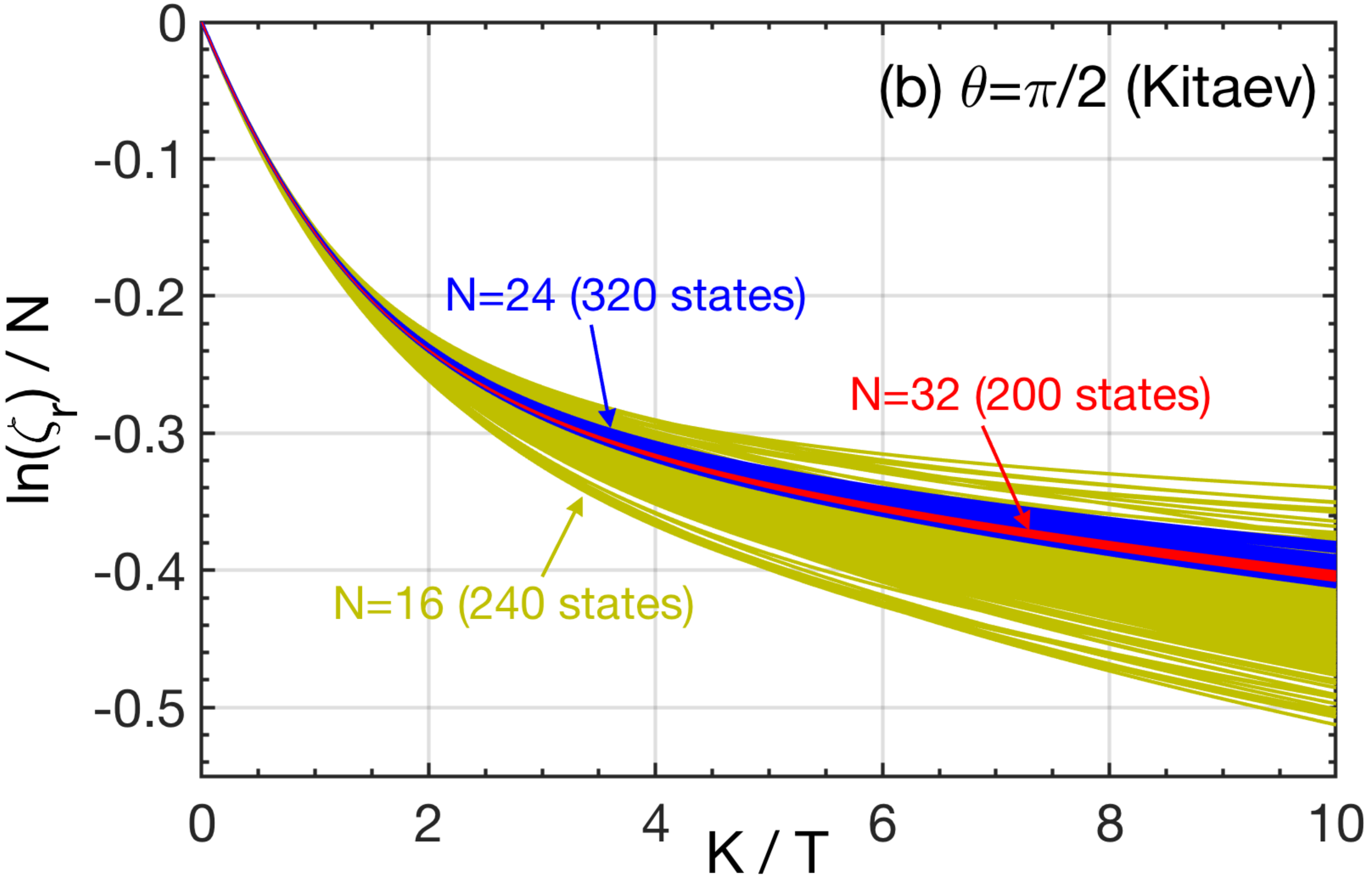}
\caption{
Evolution of the quantity $\ln(\zeta_r)/N$, where $\zeta_r\!=\!\langle r,\beta|r,\beta\rangle$, with inverse temperature for a number of initial random states $|r\rangle$ (same as in Fig.~\ref{fig:EoN}).  Panels (a) and (b) correspond to the AF Heisenberg and the AF Kitaev point, respectively.
}\label{fig:zetar}
\end{figure}

We can test this numerically by checking when the quantity $\ln(\zeta_r)/N$ is independent of system size and $r$. 
This is demonstrated in Fig.~\ref{fig:zetar}, which shows this quantity for the same clusters and number of initial states considered in Fig.~\ref{fig:EoN}.  For the N\'eel state [Fig.~\ref{fig:zetar}\,{\bf a}], the curves begin to deviate from each other around $J/T\!\sim\!1$, which roughly coincides with the coupling at which the quantities $\langle\mc{H}\rangle_{r,\beta}$ begin to converge to the ground state energy of the given sector (compare with Fig.~\ref{fig:EoN}). 

For the Kitaev point [Fig.~\ref{fig:zetar}\,{\bf b}], on the other hand, the various curves for the quantity $\ln(\zeta_r)/N$ converge to each other, and the spreading of the distribution becomes again narrower and narrower as we increase the system size. 
This picture is in essence a numerical proof of Eq.~(\ref{eq:zetar}) and the statement that the density of states is exponentially large with system size, at least for the $24$- and $32$-site clusters. 
Equation (\ref{eq:zetar}) also says that most of the weight of $|r,\beta\rangle$ comes from eigenstates within the window around $E^\ast$, which is the first ingredient behind the success of the typicality method. 

We can proceed in a similar way to show that the second ingredient, Eq.~(\ref{eq:ii}), is also guaranteed if (\ref{eq:rho}) holds. We have
\be
\!\!\!\!\!\langle A \rangle_{r,\beta} = \frac{1}{\zeta_r} \langle r,\beta| A |r,\beta\rangle 
=\frac{1}{\zeta_r} \sum_{ii'} r_i^\ast r_{i'} e^{-\beta E_i/2}e^{-\beta E_{i'}/2} \langle i| A|i'\rangle.
\ee
As above, if the number of finite terms $r_i^\ast r_{i'}$ involved in the above sum (and in the relevant energy regime) is large enough we can replace with their mean value given in Eq.~(\ref{eq:riri}) to get
\bea
\langle r,\beta| A |r,\beta\rangle \approx 
1/D \sum_{i} e^{-\beta E_i}\langle i| A|i\rangle.
\eea
The equivalence with the canonical ensemble follows immediately and, as above, the sum over $i$ is dominated by states with $E_i\!\sim\!E^\ast$.
Altogether, the success of the typicality method boils down to the presence of a large density of states $\rho(E)$. As discussed in the main text, this is typical for finite-size spectra in the middle of the spectrum, but ceases to hold below a characteristic energy scale that depends on system size and scales with the strength of the (perturbing) interactions.
The distinctive feature of the Kitaev QSL that renders the typicality method successful down to very low $T$ is the remarkable spectral downshift (see Fig.~\ref{fig:Spectrum}), that occurs generically in highly frustrated systems.

\section{Finite-size clusters used in our simulations}\label{app:clusters}
\vspace*{-0.2cm}
Fig.~\ref{fig:Clusters} shows the four main finite-size clusters with periodic boundary conditions used in our exact diagonalizations. The one shown in the left bottom panel has 24 sites and the full point group symmetry of the infinite system. The stochastic method results and the spectra shown in the main text are taken on this cluster. The clusters shown in the left top panel and the right panel have 16 and 32 sites, respectively, and lower point groups than that of the infinite system. The results shown in Figs.~\ref{fig:EoN} and \ref{fig:zetar} are taken on the 16-, 24- and 32-site clusters. The cluster in the middle panel has 24 sites but has lower point group symmetry than the infinite system. The LTLM results of Fig.~\ref{fig:Stefanos} are taken on this cluster. 
The symmetries exploited on these clusters include the translations and real space inversion. For the $JK$-model, we have also exploited the global two-fold rotation around the ${\bf x}$-axis in spin space.


\section{Boundaries of the Kitaev Quantum Spin Liquid in the $JK$ model}\label{app:Boundaries}
\begin{figure}[!b] 
\includegraphics[width=0.45\textwidth,angle=0,clip=true,trim=0 0 0 0]{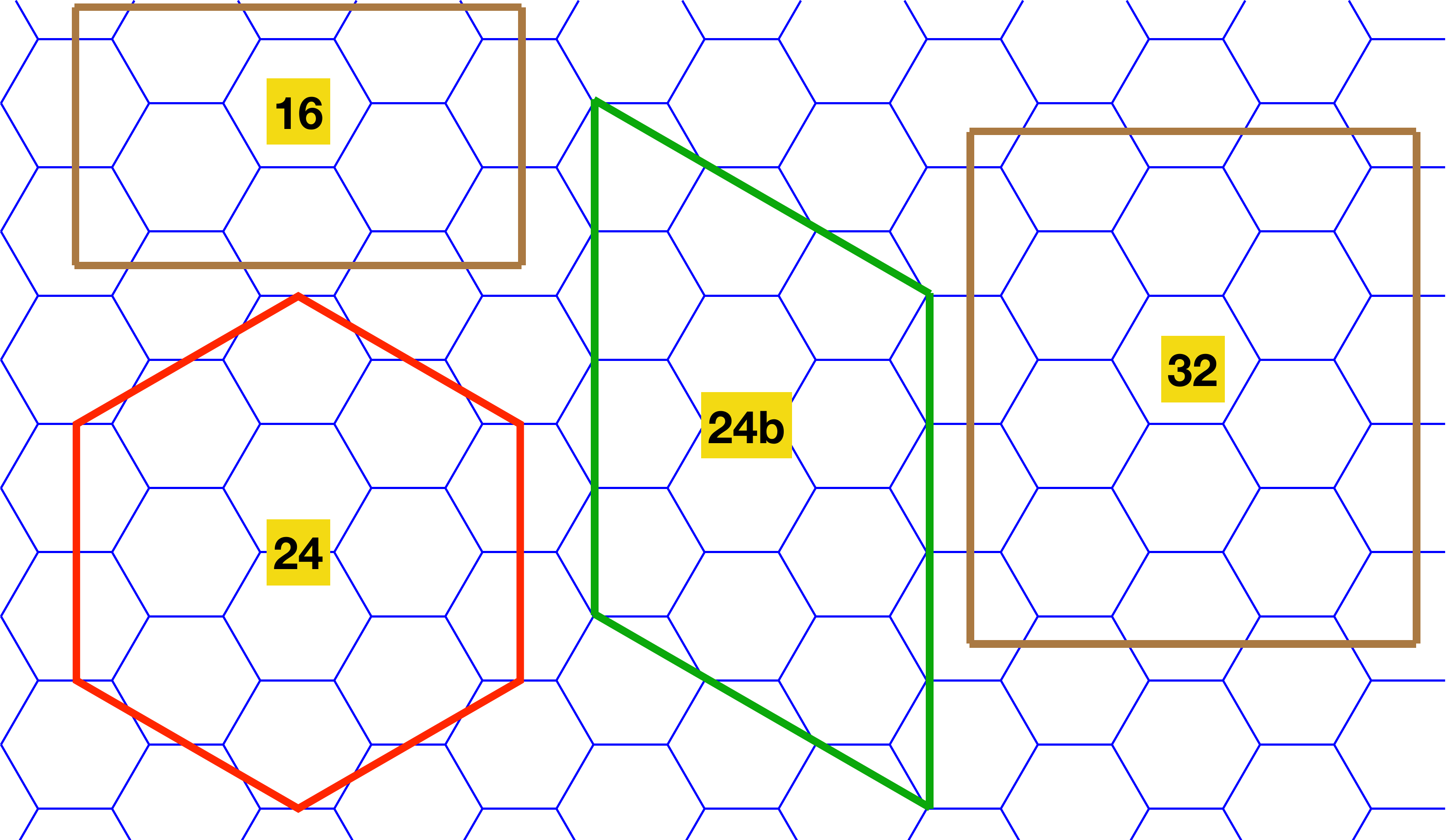}
\caption{{\bf Finite-size clusters used in our simulations.} All clusters have periodic boundary conditions, and $N\!=\!16$ (left top), $24$ (left bottom and middle) or $32$ sites (right). The first has the full point group symmetry of the model.
}\label{fig:Clusters}
\end{figure}

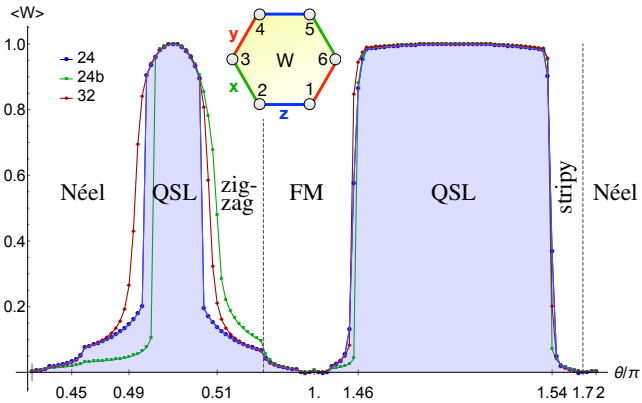
\begin{figure}[!b]
\centering\setlength{\unitlength}{\textwidth}
\begin{picture}(0.5,0.3)
\put(0,0){\includegraphics[width=0.485\textwidth,angle=0,clip=true,trim=0 0 0 0]{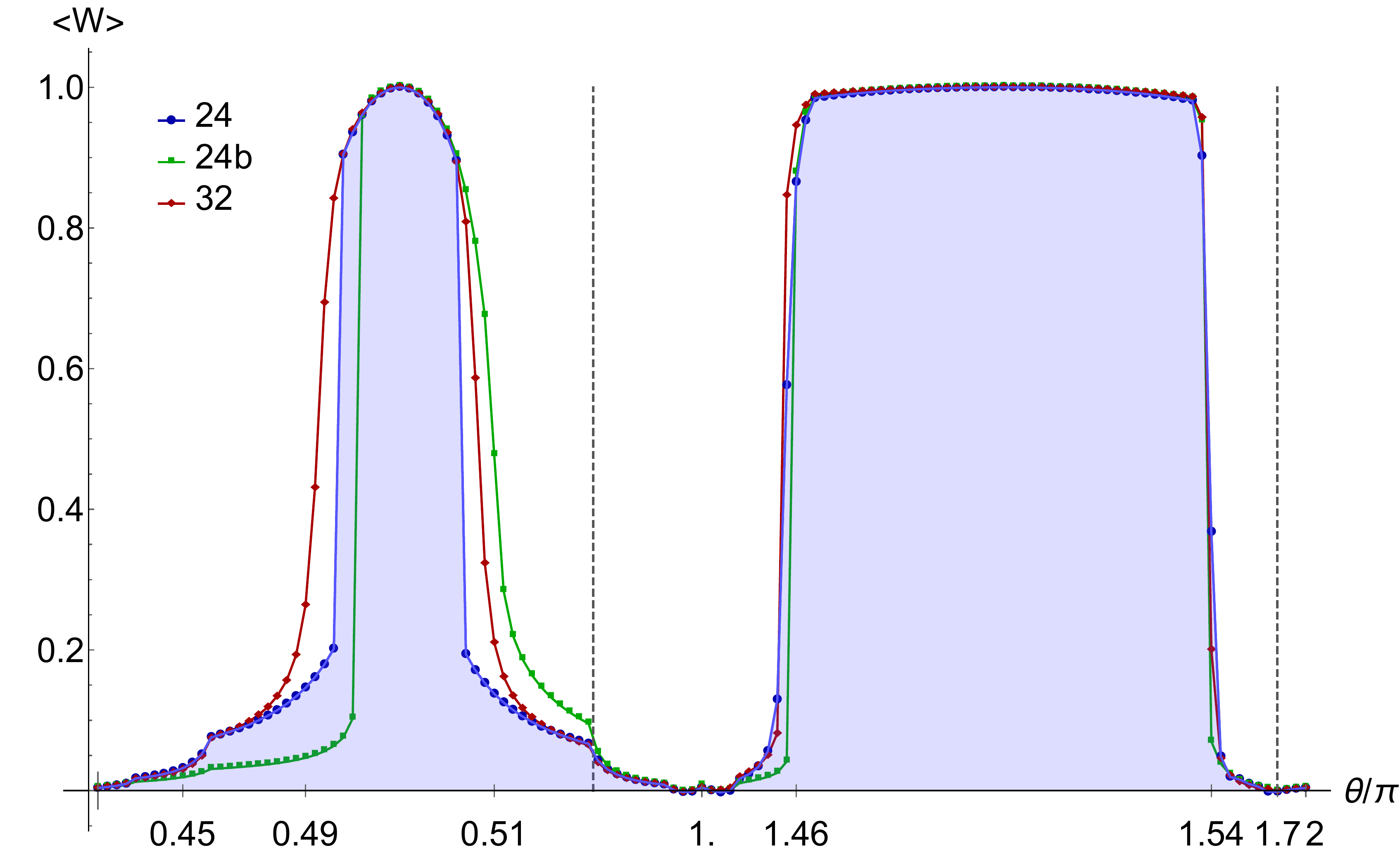}}
\put(0.178,0.21){\includegraphics[width=0.085\textwidth,angle=0,clip=true,trim=0 0 0 0]{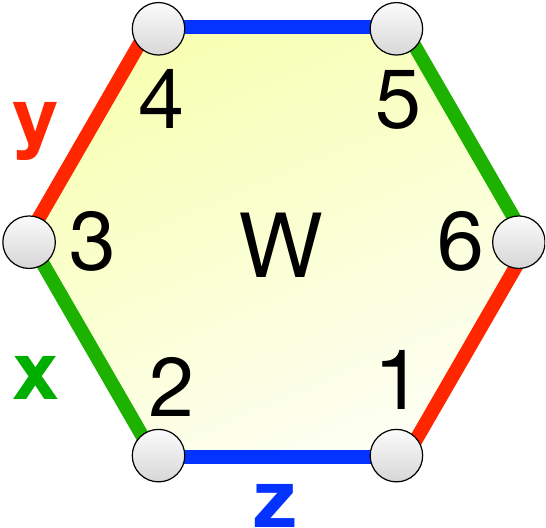}}
\put(0.055,0.15){N\'eel}
\put(0.123,0.15){QSL}
\put(0.174,0.156){zig-}
\put(0.174,0.143){zag}
\put(0.225,0.15){FM}
\put(0.33,0.15){QSL}
\put(0.422,0.14){\rotatebox{90}{stripy}}
\put(0.45,0.15){N\'eel}
\end{picture}
\caption{{\bf Phase boundaries in the $JK$-model.} Ground state expectation value of Kitaev's hexagonal plaquette operator~\cite{Kitaev2006} $W\!=\!2^6 S_1^x S_2^y S_3^z S_4^x S_5^y S_6^z$ (site labeling shown in the inset) in the $JK$-model, with $K\!=\!\sin\theta$, $J\!=\!\cos\theta$. Note the nonlinear horizontal axis. 
}\label{fig:Fluxes}
\end{figure}

\vspace*{-0.2cm}
Fig.~\ref{fig:Fluxes} shows the ground state expectation value of Kitaev's six-body flux operator~\cite{Kitaev2006} $W=2^6 S_1^x S_2^y S_3^z S_4^x S_5^y S_6^z$ as a function of $\theta$, where $K\!=\!\sin\theta$ and $J\!=\!\cos\theta$. The results are obtained from exact diagonalizations on the three clusters of Fig.~\ref{fig:Clusters}.
The Kitaev QSL phases correspond to the regions around the ideal Kitaev points $\theta\!=\!\pm\pi/2$, where $\langle W\rangle$ is very close to 1.~\cite{Kitaev2006} The transitions to the magnetically ordered phases correspond to the points where $\langle W\rangle$ drops abruptly to very small values. The shaded QSL regions in Fig.~\ref{fig:Fluxes} correspond to the symmetric 24-site cluster of Fig.~\ref{fig:Clusters} which has the full point group symmetry of the model. 
The boundaries are consistent with those reported in Ref.~[\onlinecite{Jackeli2013}] and the reorganization of the low-lying excitation spectra in Fig.~\ref{fig:Spectrum} of the main text.


\section{Results from low-$T$ Lanczos method}\label{app:LTLM}
\vspace*{-0.2cm}
The low-$T$ Lanczos method (LTLM)~\cite{Aichhorn2003} is an adaptation of the earlier, finite-$T$ Lanczos algorithm~\cite{Jaklic2000,Prelovsek2013,Schnack2018} that correctly captures the zero temperature limit. This method delivers the response directly in frequency domain and the thermodynamic trace is performed exactly at low enough $T$, by keeping contributions from all low-lying excitations below an energy cutoff. 
Specifically, one starts again with $r_m$ random states $|r\rangle$ and obtain, for each state, a set of $M$ approximate eigenstates $\{|\epsilon_i^{(r)}\rangle, i\!=\!1\text{-}M\}$ by a standard Lanczos iteration scheme.~\cite{Lanczos1950,Paige1971,CullumWilloughby1,Saad2011}  
The partition function is then approximated by
\be
Z \!\simeq\! \frac{D}{r_m} \sum_{r=1}^{r_m} \sum_{i=1}^M |\braket{\varepsilon_i^{(r)}|r}|^2 e^{-\beta\varepsilon_i^{(r)}} \,,
\ee
and any given spectral density $C(\omega)\!\equiv\!\frac1\pi\!\int\!dt e^{i\omega t} \langle A^\dagger(t) A(0)\rangle$ by
\be\label{eq:ltlmraman}
\begin{array}{c}
C(\omega) \simeq\frac{D}{Z r_m} \sum_{r=1}^{r_m} \sum_{i,l,k=1}^M e^{-\beta (\varepsilon_i^{(r)} + \varepsilon_k^{(r)})/2} \braket{r|\varepsilon_i^{(r)}}  \braket{\varepsilon_i^{(r)}|A^\dagger|\tilde\varepsilon_l^{(r)}} \\ 
\\
~~~~~~ \times  \braket{\tilde\varepsilon_l^{(r)}|A|\varepsilon_k^{(r)}} \braket{\varepsilon_k^{(r)}|r} ~\delta\left(\omega - \tilde\varepsilon_l^{(r)} + \frac12 (\varepsilon_i^{(r)} + \varepsilon_k^{(r)})\right) \,.
\end{array}
\ee
Here, $D$ is the dimension of the full Hilbert space and $\ket{\tilde\varepsilon_l^{(r)}}$ are eigenstates of $\mc{H}$ with corresponding eigenenergies $\tilde\varepsilon_l^{(r)}$, obtained by an additional Lanczos run with initial state $A \ket{\varepsilon_k^{(r)}}$ for each $k$.
In practice, $M\!\sim\!\mc{O}(100)$ is sufficient to obtain accurate estimates for observables of a sparse matrix $\mc{H}$, even for $D\!\sim\!\mc{O}(10^8)$. Nevertheless, the computational cost is prohibitive if all three sums are over Lanczos bases with $M\! \sim\!100$. Since higher-energy eigenvalues are exponentially suppressed by their Boltzmann factors at low $T$, one can further restrict the summations over $i$ and $k$, so that $\varepsilon_{i,k} < \varepsilon_c$ and the energy threshold $\varepsilon_c$ is in turn set by the condition
\begin{equation}
 e^{-\beta_c (\varepsilon_c - \varepsilon_0)} < \epsilon_c \,,
\end{equation}
where $\epsilon_c$ is the high-energy fraction of the Boltzmann factors to be truncated and $\beta_c$ is the lowest inverse temperature one is interested in. For example, $\epsilon_c\!=\!0.05$ means that one ignores all eigenstates that contribute 5\% of the high-energy tail of the Boltzmann weight. A good accuracy is thus guaranteed for all $\beta\!\geq\!\beta_c$.
Furthermore, the calculation can be performed independently in each symmetry sector (as in the stochastic method), and the final expectation value is then obtained after an additional averaging over all sectors.
Finally, the $\delta$-peaks in Eq.~(\ref{eq:ltlmraman}) are broadened into Lorentzians by a common broadening parameter $\eta$. 
Here, the value of $\eta$ has been chosen to facilitate a comparison with results from the typicality method, which misses the finer finite-size structure of the spectra at very low $T$ (due to the breakdown of typicality below $T_{\text{typ}}$).

\begin{figure}[!t] 
\includegraphics[width=0.42\textwidth,angle=0,clip=true,trim=0 0 0 0]{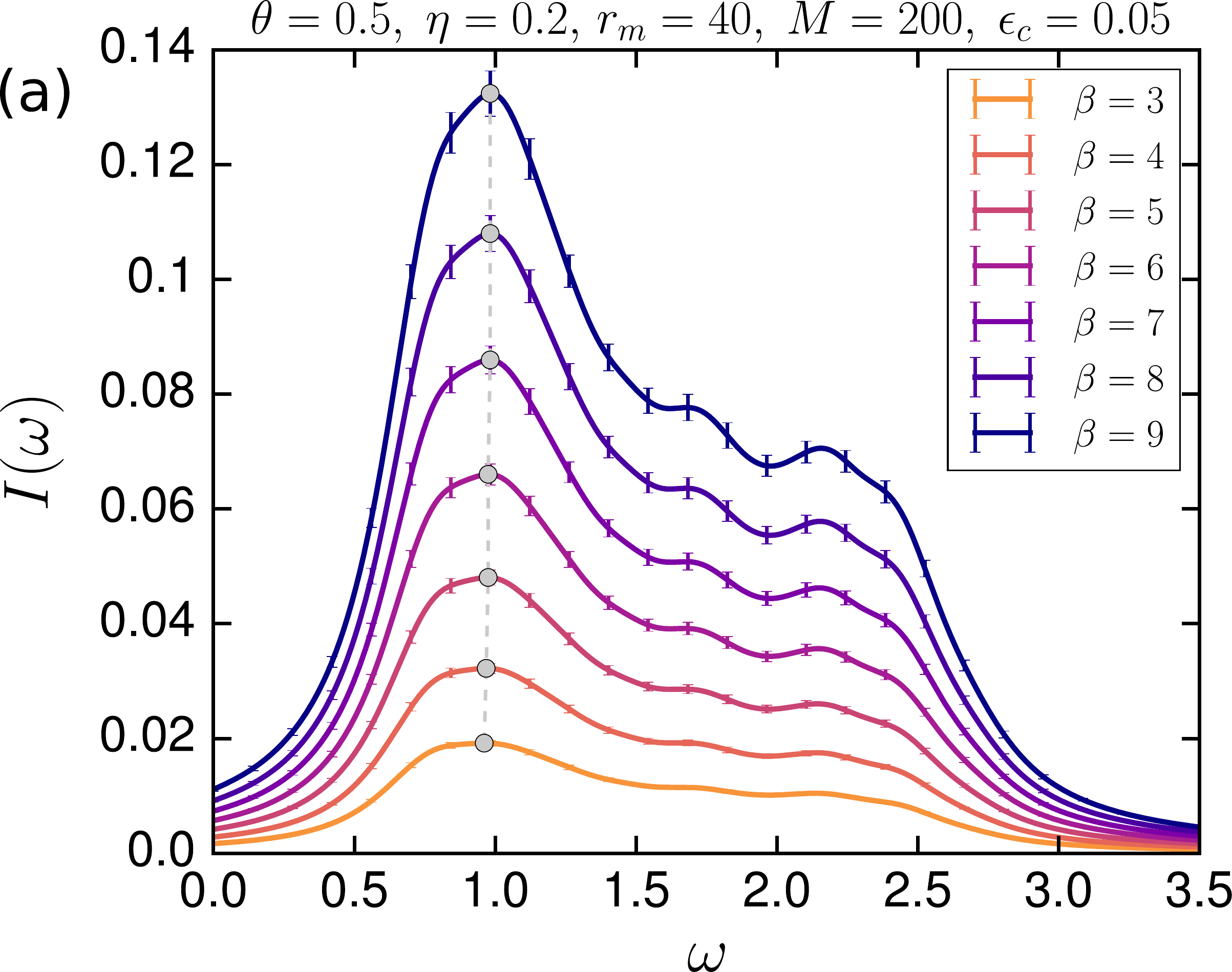}
\vspace{1em}
\includegraphics[width=0.42\textwidth,angle=0,clip=true,trim=0 0 0 0]{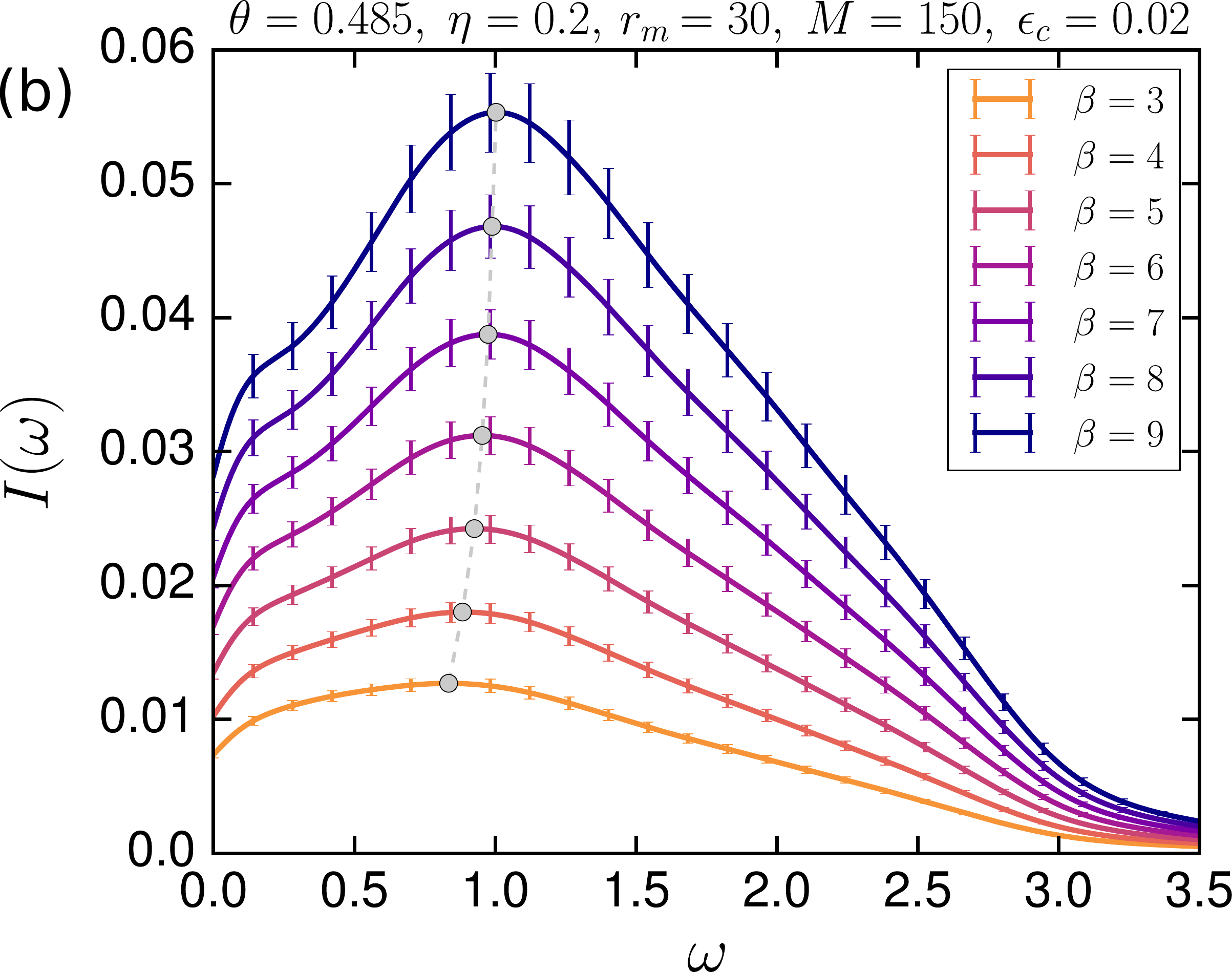}
\caption{{\bf Results for the LTLM method.} Raman response in the $xy$-channel at ({\bf a}) the Kitaev QSL point, and ({\bf b}) $\theta\!=\!0.485\pi$ (within the N\'{e}el phase). Standard error due to the sampling involved in Eq.~\eqref{eq:ltlmraman} is shown as error bars. Dashed grey lines mark the shift of the main peak with $T$. Peaks are broadened with a Lorentzian of width $\eta\!=\!0.2$.
}
\label{fig:Stefanos}
\end{figure}

We have used the LTLM to study the $xy$-channel Raman response of the Kitaev-Heisenberg model for periodic clusters of up to 24 sites in the window $\beta \geq \beta_c\!=\!3$, within which numerical calculations can be performed with available resources. 
Two representative set of results are shown in Fig.~\ref{fig:Stefanos} for the cluster labeled as `24b' in Fig.~\ref{fig:Clusters}, along with the values of the LTLM parameters $r_m$, $M$, $\eta$ and $\epsilon_c$ used. We have verified that the results do not change appreciably when these parameters are varied.  
The spectra share the same qualitative features with those from the typicality method. Finer differences arise due to the different point group symmetries of the clusters `24b' and `24' of Fig.~\ref{fig:Clusters}. Close to the Kitaev point, but on either side of its boundary with the N\'{e}el phase, the Raman response has two main features, namely, a primary peak at $\omega\!\sim\!1$ and a secondary peak/shoulder at $\omega\!\sim\!2$, which disperse with increasing $T$ to lower and higher frequencies respectively (more markedly for $\theta\!=\!0.485\pi$). These observations are also in accordance with published Monte Carlo results at the Kitaev point.~\cite{Nasu2016}
\\


%

\end{document}